\pdfoutput=1
\documentclass[11pt]{article}

\usepackage[preprint]{jmlr2e}

\usepackage{microtype}
\usepackage{amsmath}
\usepackage{multicol}
\usepackage{algorithmic}
\usepackage[all]{xy}
\usepackage{placeins}

\DeclareMathOperator*{\argmax}{argmax}

\ShortHeadings{Neural-prior SBM}{Neural-prior SBM}
\firstpageno{1}

\begin{document}

\title{Neural-prior stochastic block model}

\author{\name \includegraphics[height=0.62em]{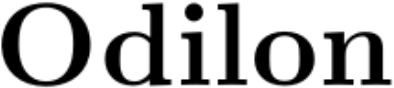} Duranthon \\
       \addr Statistical Physics of Computation laboratory (SPOC)\\
       EPFL, Lausanne, Switzerland
       \AND
       \name Lenka Zdeborov\'a \\
       \addr Statistical Physics of Computation laboratory (SPOC)\\
       EPFL, Lausanne, Switzerland   }

\editor{}

\maketitle

\begin{abstract}

The stochastic block model (SBM) is widely studied as a benchmark for graph clustering aka community detection. In practice, graph data often come with node attributes that bear additional information about the communities. Previous works modeled such data by considering that the node attributes are generated from the node community memberships. In this work, motivated by a recent surge of works in signal processing using deep neural networks as priors, we propose to model the communities as being determined by the node attributes rather than the opposite. We define the corresponding model; we call it the neural-prior SBM. We propose an algorithm, stemming from statistical physics, based on a combination of belief propagation and approximate message passing. We analyze the performance of the algorithm as well as the Bayes-optimal performance. We identify detectability and exact recovery phase transitions, as well as an algorithmically hard region. The proposed model and algorithm can be used as a benchmark for both theory and algorithms. To illustrate this, we compare the optimal performances to the performance of simple graph neural networks.

\end{abstract}


\section{Introduction}

The stochastic block model (SBM) is widely studied as a benchmark for graph clustering aka community detection, see e.g. reviews \cite{fortunato2010community,abbe2017community,peixoto2019bayesian}. The standard version of the stochastic block model observed a graph of connections and the goal is to recover the communities from the knowledge of the graph. 

However, in practice, graph data often come with node attributes that bear additional information about the communities. In such a case there are several sources of information on communities one can use: the structure of the graph (as in the standard SBM), and the features or attributes of the nodes. Past work developed algorithms and models accounting for such node information. Among the well-known is the CESNA model of \cite{yang2013community} where the attributes are generated via logistic regression on the community membership. Another model that recently became popular in the context of benchmarking graph neural networks (e.g. \cite{benchmarkingCSBM18,attentionCSBM22,benchmarkingCSBM22}) is the contextual SBM \cite{cSBM14,cSBM18}, where communities determine centroids for a Gaussian mixture model generating the node-features. In both these examples, the node attributes are generated via conditioning on the community label of the node.

In signal processing, another separate line of work, that witnesses a surge of interest, is modeling signals as the output of a deep generative neural network; for recent reviews see e.g. \cite{ongie2020deep,shlezinger2020model}. Deep generative neural networks can be trained on data, and due to their expressivity are able to capture generic structural properties of the signal. In community detection the signal can be seen as the community memberships; following the line of work on deep generative priors it is hence of interest to propose a model where the node attributes are an input of a generative neural network and the node community memberships are the output thereof. In this work, motivated by a recent surge of works in signal processing using deep neural networks as priors, we propose to model the communities as being determined by the node attributes rather than the opposite. We define the corresponding model; that we call the neural-prior SBM.

One of the attires of the stochastic block model is that it is amenable to exact statistical analysis of what is the best achievable performance from an information-theoretic and from algorithmic point of view. This has led to a line of work, originating in statistical physics, where statistical and computational thresholds are analyzed; see e.g. \cite{SBM11,abbe2015exact,abbe2017community}. It is valuable to have a solvable case for which we know what is statistically and algorithmically achievable; because in the context of modern machine learning, it is rarely known if or how much the observed performance can be further improved. Asymptotically exact analysis of the detectability threshold was also performed for the contextual stochastic block model \cite{cSBM18,lu2020contextual}. The main topic of the present paper is the statistical physics analysis of optimal algorithmic performance for a simplified version of the proposed neural-prior stochastic block model that we call the generalized-linear-model SBM (GLM--SBM).

The GLM--SBM model we propose can be used for benchmarking graph neural networks (GNNs). Since the model is analyzable, we can compare the performance of the evaluated GNN to the optimal algorithmic performance in a non-trivial high-dimensional setting. We treat both the unsupervised and the semi-supervised cases and accompany our paper with an implementation that can be readily used for comparison by GNN developers. As far as we found, a model similar to the neural-prior SBM, we propose here, has been used in \cite{transformerSBM22}. In that work it is used as a building block for a large neural network; it was not analyzed per se.

A large part of this paper is dedicated to the asymptotic analysis of the GLM--SBM model. We identify how the detectability phase transition well known from the SBM changed under the presence of the GLM-prior. We also unveil an exact recovery phase transition that happens when the prior on the latent variables of the GLM is binary, while the average degree of the SBM remains finite. Such an exact recovery phase at a finite average degree came to us as a surprise and we find it rather remarkable in view of the fact that without the GLM prior exact recovery in the standard SBM is only possible for degrees growing logarithmically with the system size \cite{abbe2015exact,abbe2017community}. The exact recovery transition is discontinuous and makes the problem algorithmically challenging posing a nice set of parameters that can serve as a benchmark in the attempt of improving graph-neural networks. 

\section{The neural-prior stochastic block model}

\subsection{Definition}
We consider a set $V$ of $|V|=N$ nodes, a graph $G(V,A)$ on those nodes. Nodes have features/attributes $F_\mu \in \mathbb R^M$ of dimension $M$, $\mu = 1 \dots, N$. The features and the graphs are observed. We aim to divide the set of nodes into $q$ communities with labels $s_\mu \in \{1,\dots,q\}$ in such a way that (a) the graph structure correlates with the labels, e.g. nodes being in the same community are more likely to be connected, and (b) the node attributes $F_\mu$ are correlated with the labels. 

\paragraph{SBM:} In the stochastic block model the edges $A_{\mu\nu}$ of the graph $G$ are generated conditioned on the group memberships $s_\mu$; we consider the following rule:
\begin{equation}
P_{\rm SBM}(A_{\mu\nu}=1|s_\mu,s_\nu) =
\left \{
\begin{array}{r c l}
  c_i/N & \mathrm{if} & s_\mu=s_\nu, \\
  c_o/N & \mathrm{if} & s_\mu \neq s_\nu,
\end{array}
\right .
\end{equation}
and $A_{\mu\nu}=0$ otherwise. Here $c_i$ and $c_o$ are the affinity coefficients common to the SBM. We define the affinity matrix whose elements are $c_{s,t}=c_i\delta_{s=t}+c_o\delta_{s\neq t}$. 
We note that the literature often considers a more general SBM where the affinity matrix has arbitrary elements. The model and analysis proposed in this work could be readily generalized to that case. We consider a slightly restricted version of the SBM purely for simplicity.
In the SBM the ground truth group memberships $s_\mu$ are generated at random from a prior that only accounts for the sizes of the $q$ groups. The node attributes $F$ are simply ignored in the SBM. 

\paragraph{Neural-prior SBM:} In neural-prior SBM, that we define here, the group memberships $s_\mu$ can be a generic function on the attributes $F_\mu$. Such a function can be represented by a deep neural network and learned from ground-truth data. The training data would be pairs $\{F_\mu,s_\mu\}$ where attributes act as the neural network inputs and the group memberships as output labels. For instance, for a $L$-layer fully connected neural network this reads 
\begin{equation}
\label{eq:multi}
s_\mu=\varphi^{(L)} \big(W^{(L)}  \dots \varphi^{(2)} \big( W^{(2)} \varphi^{(1)} ( W^{(1)} F )) \dots)
\end{equation}
for the last activation function $\varphi^{(L)}$ chosen as in multi-class classification tasks. 

The aim of this paper is to provide a benchmark model where the optimal performance can be analyzed asymptotically exactly. For this we need to (a) define the corresponding asymptotic limit, (b) consider a simple neural network prior that is amenable to asymptotic analysis. We will also limit ourselves to consider community detection with two groups of the same size only, $q=2$ (this is not a strong limitation, but is considered in the follow-up for simplicity). With this in mind, in the rest of the paper, we will consider the following model generating the group memberships $s_\mu$.  

\paragraph{GLM--SBM:} In order to make analysis amenable we will consider the features $F$ to be random and drawn independently as $F_{\mu l} \sim\mathcal N(0,1/M)$. We then consider $M$ latent variables $w_l \sim P_w$, $l=1,\dots,M$ and generate the community memberships as 
\begin{equation}
s_\mu=\mathrm{sign}\big( \sum_l^M F_{\mu l}w_l \big)
\end{equation}
This corresponds to a single-layer neural network with a sign activation function. Such a neural network is also often referred to as the generalized linear model (GLM) or as the perceptron. We will hence call this variant of the neural-prior SBM the GLM--SBM. 

Concerning the asymptotic limit, we work in the challenging sparse case of SBM. We parameterize the SBM by the standard parameterization
\begin{equation}
c_i = c+\sqrt c\lambda \quad,\quad c_o = c-\sqrt c\lambda
\end{equation}
We then consider $N\to \infty$ with $c=(c_i+c_o)/2 = \mathcal O(1)$ is the average degree, and $\lambda = \mathcal O(1)$ is the signal-to-noise ratio. We further work in the high-dimensional limit of the GLM where $N/M=\alpha= \mathcal O(1)$, with $\alpha$ being the aspect ratio that will play a role of another signal-to-noise ratio. This is because the higher $\alpha$ the more correlation there is between the group memberships and the easier the community detection should be. 

The GLM--SBM differs from the SBM because communities are not independent, conditionally on the features. For instance, in the extreme case $M=1$, all memberships are known, up to a global flip given by $w_1$; that is to say, they are all very strongly correlated. The GLM--SBM tends toward a SBM when $\alpha\to 0$. Indeed, for large $M$, $\sum_l^M F_{\mu l}w_l$ tend to independent Gaussian variables.

\subsection{Related work}

Anticipating the asymptotic analysis that we are aiming at, we note that such an analysis has been done for the standard SBM in \cite{SBM11,decelle2011inference} using the belief propagation algorithm and the cavity method from statistical physics for the asymptotic analysis of its behavior. Concerning semi-supervised learning in the SBM, the information coming from the semi-supervision is readily incorporated into the analysis of the above papers as has been done in \cite{zhang2014phase}. 

The predictions of \cite{SBM11,decelle2011inference} have then been partially established rigorously see e.g. \cite{mossel2015reconstruction,mossel2018proof,abbe2017community,coja2017information}. However, the full conjecture of \cite{SBM11,decelle2011inference} about the asymptotic exactness of their analysis remains an open question from the mathematical point of view. In this paper, we will use the same techniques as  \cite{SBM11,decelle2011inference} anticipating a follow-up work putting the conjectures about optimality on a rigorous basis. For the GLM, which is defined by a dense graphical model, the rigorous analysis is simpler and was carried out in \cite{barbier2019optimal}. 

The analysis of the GLM--SBM requires to \emph{glue} the two graphical models using the GLM as the prior for the SBM, and the SBM as a source of uncertainty of the outputs of the GLM. Such a glueing of two dense exactly solvable graphical models for developed in \cite{manoel2017multi} with rigorous justifications given in \cite{gabrie2018entropy,aPrioriGen19,gerbelot2021graph}. Our work is the first one, as far as we are aware, where a sparse graphical model (the SBM) is glued to a dense graphical model (the GLM). This can be done heuristically and is conjectured asymptotically exact along the lines of the works of \cite{SBM11,decelle2011inference}. A complete rigorous justification would have to be preceded by the proof of the conjecture for the SBM that is still open. 

The contextual stochastic block model (CSBM) introduced and studied theoretically in \cite{cSBM14,cSBM18} is another version of the SBM incorporating node information. In the CSBM the node information is modeled via a Gaussian mixture model with each community having their own centroid. From the analysis point of view, this model takes into account two sources of observation about the latent variables -- the community memberships. This is hence different from the GLM--SBM where one model serves as a prior for the other instead of as an independent source of information. Modulo this difference, some of the analysis performed for the CSBM is related to our work. Notably, the detectability threshold and the linearized message passing algorithm presented in \cite{cSBM18,lu2020contextual} are obtained in a similar manner in which we obtain the detectability phase transition and the linearized algorithm. We note that the semi-supervised version of the CSBM has not been analyzed, but this could be done rather straightforwardly using the same methods as in \cite{zhang2014phase}.

\section{Bayes-optimal estimation of communities}

We consider the GLM--SBM as defined above and aim to analyze the Bayes-optimal inference of the community structure. We will consider in general the semi-supervised setting where next to the structure of the graph $A$ and the covariates $F$ we observe the communities for a subset $\Xi$ of the nodes, $\rho = |\Xi|/N$. We denote by $s$ the vector of unobserved nodes and $s_\Xi$ the vector of observed nodes. The unsupervised case is then recovered as the special case where $\Xi$ is an empty set, $\rho=0$.

The analysis of this paper is set in the so-called Bayes-optimal setting where we know the details of the GLM--SBM model. The only quantity that we do not observe is the ground truth values of the latent variables $w$ that generate the group memberships $s$. For the group memberships, we only observe a fraction $\rho$ of them in the semi-supervised setting and none of them in the unsupervised setting. 

The optimal inference is then done using the posterior distribution over the unobserved communities 
\begin{equation}
P(s|A,s_\Xi,F) = \frac{P(A|s,s_\Xi,F)P_{\rm prior}(s|s_\Xi,F)}{Z(A,s_\Xi,F)} = \frac{P_{\rm prior}(s|s_\Xi,F)}{Z(A,s_\Xi,F)}\prod_{\mu<\nu}P_{\rm SBM}(A_{\mu\nu}|s_\mu,s_\nu)
\end{equation}
where $Z(A,F,s_\Xi)$ is the normalization constant. We used here the definition of the GLM--SBM model that implies $P(A|s,s_\Xi,F)=P(A|s,s_\Xi)$. For GLM--SBM the prior on $s$ is
\begin{equation}
\label{eq:posterior_s}
P_{\rm prior}(s|s_\Xi,F) = \frac{1}{Z(s_\Xi,F)}\int\mathrm dw\,P_w(w)\prod_\mu \Big[ P_{s,\mu}(s_\mu)P_0(s_\mu|\sum_l^MF_{\mu l}w_l)\Big]
\end{equation}
where we define $P_0(t|z)=\delta_{t=\mathrm{sign}(z)}$ the output distribution and $P_{s,\mu}$ the additional prior distribution, which is used to inject information about the membership of node $\mu$:
\begin{equation}
P_{s,\mu}(t)=
\left \{
\begin{array}{r c l}
  \delta_{t=s_\mu} & \mathrm{if} & \mu\in\Xi, \\
  1/2 & \mathrm{if} & \mu\notin\Xi.
\end{array}
\right .
\end{equation}

In eq.~\eqref{eq:posterior_s} we marginalize over the latent variable $w$. However, since the estimation of the latent variable $w$ is crucial in order to exploit the full power of the prior~\eqref{eq:posterior_s} it will be instrumental to consider the posterior as a joint probability of the unobserved nodes and the latent variable 
\begin{equation}
\label{eq:posterior_s_w}
P(s,w|A,s_\Xi,F) = \frac{  P_w(w)}{\tilde Z(A,s_\Xi,F)}  \prod_\mu \Big[P_{s,\mu}(s_\mu)P_0(s_\mu|\sum_l^MF_{\mu l}w_l)\Big] \prod_{\mu<\nu}P_{\rm SBM}(A_{\mu\nu}|s_\mu,s_\nu)
\end{equation}
$\tilde Z$ is the Bayesian evidence. We define the free entropy of the problem as its logarithm: 
\begin{equation}
 \phi(A,s_{\Xi},F) = \frac{1}{N}\log \tilde Z(A,s_{\Xi},F)
\end{equation}

We seek an estimator $\hat s$ that maximizes the overlap with the ground truth. The Bayes-optimal estimator $\hat s$ that maximizes it is given by
\begin{equation}
\label{eq:estimateurMMO}
\hat s_\mu^\mathrm{MMO}=\argmax_t\, p_\mu(t)
\end{equation}
where $p_\mu$ is the marginal posterior probability of node $\mu$. Using the ground truth values $s_\mu$ of the communities the maximal mean overlap is then computed as
\begin{equation}
\mathrm{MMO}=\frac{1}{N}  \sum_{\mu=1}^N \delta_{\hat s_\mu^\mathrm{MMO}, s_\mu}
\end{equation}
To estimating the latent variable $w$, we consider minimizing the mean squared error via the MMSE estimator 
\begin{equation}
\label{eq:estimateurMMSE}
\hat w_l^\mathrm{MMSE}=\int\mathrm ds\mathrm dw\,P(s,w|A,s_{\Xi},F)w_l
\end{equation}
i.e. $\hat w^\mathrm{MMSE}$ is the mean of the posterior distribution. Again using the ground truth values $w_l$ of the latent variables the MMSE is then computed as 
\begin{equation}
\mathrm{MMSE}=\frac{1}{M}\sum_{l=1}^M(\hat w_l^\mathrm{MMSE}-w_l)^2
\end{equation}
The problem is invariant by a global sign flip of $s$ and $w$ so in practice we measure the following overlaps
\begin{equation}
q_S = \frac{|\hat s \cdot s|}{N} \quad,\quad q_W = \frac{|\hat w \cdot w|}{||\hat w||_2 ||w||_2}
\end{equation}

In general, the Bayes-optimal estimation requires the evaluation of the averages over the posterior that is in general exponentially costly in $N$ and $M$. In the next section, we will derive the AMP--BP algorithm and argue that, in the limit $N\to \infty$ and $M\to \infty$ with $N/M= \alpha = {\cal O}(1)$ and all other parameters being of ${\cal O}(1)$ this algorithm approximates the MMSE and MMO estimators with an error that vanishes. We give more precise statements below.

\section{The AMP--BP algorithm}

To retrieve the communities for the GLM--SBM, our main results rely on an algorithm that we call AMP--BP. We conjecture that in the large system size, this algorithm cannot be beaten by another polynomial algorithm. We can also extract the so-called hard phases where the randomly initialized algorithm fails, but an exponentially costly algorithm would succeed; we do this using an informed initialization and the free entropy. We then analyze the performance of the algorithm and the associated phase transitions.  

\subsection{Algorithm}
The algorithm is based on belief propagation (BP) and approximate message-passing (AMP). BP was used to solve SBM in \cite{SBM11} and conjectured asymptotically optimal among efficient algorithms in doing so. AMP was used to solve GLM, see e.g. \cite{donoho2009message,GLM12}, and again conjectured asymptotically optimal among efficient algorithms in doing so with strong evidence for this being provided by \cite{celentano2021high}. We glue these two algorithms together along the lines of \cite{manoel2017multi,aPrioriGen19} to solve the GLM--SBM; we call the resulting algorithm AMP--BP. Using statistical physics arguments analogous to those in \cite{SBM11,GLM12} we conjecture that it provides asymptotically optimal performance in the considered cases.

We derive the AMP--BP algorithm for the GLM--SBM starting from the factor graph of the problem:\\
\centerline{
\xymatrix{
 *+[o][F-]{w_l} \ar@{-}[r]\ar@{-}[dr] \ar@{}[r]^(.25){}="a"^(.75){}="b" \ar@<3pt>"a";"b"^{\chi_w^{l\to\mu}} & *+[F]{ } \ar@{-}[r] \ar@{}[r]^(.25){}="a"^(.75){}="b" \ar@<3pt>"a";"b"^{\psi_s^{\mu\to\mu}} & *+[o][F-]{s_\mu} & {}\save[]+<0cm,-0.65cm>*+[F]{ }  \ar@{-}[l] \ar@{}[l]^(.25){}="a"^(.75){}="b" \ar@<3pt>"b";"a"^{\chi_s^{\mu\to\nu}} \ar@{-}[ld] \ar@{}[ld]^(.25){}="a"^(.75){}="b" \ar@<3pt>"a";"b"^{\psi_s^{\mu\to\nu}} \restore \\
 *+[o][F-]{w_m} \ar@{-}[r]\ar@{-}[ur] \ar@{}[r]^(.25){}="a"^(.75){}="b" \ar@<3pt>"b";"a"^{\psi_w^{\nu\to m}} & *+[F]{ } \ar@{-}[r] \ar@{}[r]^(.25){}="a"^(.75){}="b" \ar@<3pt>"b";"a"^{\chi_s^{\nu\to\nu}} & *+[o][F-]{s_\nu} &
}
}
The $\chi$s and $\psi$s are probability distributions on the variables $s$ and $w$; they are called cavity messages. We write the belief-propagation (BP) equations for these distributions that read:  
\begin{align}
\chi_{w_l}^{l\to\mu} &\propto P_w(w_l)\prod_{\nu\neq\mu} \psi_{w_l}^{\nu \to l}\\
\psi_{w_l}^{\nu\to l} &\propto \sum_{s_\nu}\chi_{s_\nu}^{\nu \to\nu} \int\prod_{m\neq l}\left(\mathrm dw_m \chi_{w_m}^{m\to\nu}\right) P_0(s_\nu | F_\nu\cdot w)\\
\psi_{s_\nu}^{\nu\to\nu} &\propto \int \prod_m\left(\mathrm dw_m \chi_{w_m}^{m \to\nu}\right) P_0(s_\nu | F_\nu\cdot w)\\
\chi_{s_\mu}^{\mu\to\mu} &\propto P_{s,\mu}(s_\mu)\prod_{\nu\neq\mu} \psi_{s_\mu}^{\nu\to\mu}\\
\chi_{s_\mu}^{\mu\to\nu} &\propto P_{s,\mu}(s_\mu)\psi_{s_\mu}^{\mu\to\mu}\prod_{\eta\neq\mu,\nu} \psi_{s_\mu}^{\eta\to\mu}\\
\psi_{s_\nu}^{\mu\to\nu} &\propto \sum_{s_\mu}\chi_{s_\mu}^{\mu\to\nu} P_{\rm SBM}(A_{\mu\nu} | s_\mu,s_\nu)
\end{align}
The proportionality signs $\propto$ denote that all the messages are non-negative numbers summing to one over their lower indices, the corresponding normalization factors being omitted in our notation. 

These BP equations still include a high-dimensional integral and hence cannot be implemented efficiently. We simplify them to obtain AMP--BP by using the central limit theorem on the dense side of the graphical model and keeping only the means and variances of the resulting Gaussians. This is standard in the derivation of the AMP algorithm, see e.g. \cite{GLM12}. The details of this derivation are given in appendix \ref{derivationAlgo}.

In order to state the final algorithm, we introduce the denoising function:
\begin{equation}
g_o(\omega, \chi, V) = \frac{\int\mathrm dz\sum_s\chi_s P_0(s|z)(z-\omega)e^{-(z-\omega)^2/2V}}{V\int\mathrm dz\sum_s\chi_s P_0(s|z)e^{-(z-\omega)^2/2V}}
\end{equation}
We define the input functions as
\begin{equation}f_a(\Lambda, \Gamma) = \frac{\int\mathrm dw\,P_w(w)we^{-\Lambda w^2/2+\Gamma w}}{\int\mathrm dwP_w(w)e^{-\Lambda w^2/2+\Gamma w}} \quad,\quad f_v(\Lambda, \Gamma)=\partial_\Gamma f_a(\Lambda, \Gamma)
\end{equation}
We denote by $Z$s the normalization factors obtained so that the messages sum to one over their lower indices.

To give some intuition we explain what are the variables AMP--BP employs. $a_l$ is an estimation of the posterior mean of $w_l$, $v_l$ is an estimation of its variance; $\omega_\mu$ is an estimation of the mean of $\sum_l F_{\mu l}w_l$ and $V$ an estimation of its variance. $\psi_{s_\mu}^{\mu\to\mu}$ is a marginal distribution on $s_\mu$, as estimated by the AMP on the GLM side, while $\chi_{s_\mu}^{\mu\to\mu}$ is the distribution as estimated by the BP on the SBM side. $\Gamma_l$ is a proxy for estimating the mean of $w_l$ in absence of the prior $P_w$ and $\Lambda$ is for the variance. $h_t$ can be interpreted as an external field enforcing the nodes not to be in the same group; $\chi_{s_\mu}^{\mu\to\nu}$ is a marginal distribution on $s_\mu$ (these variables are the messages of a sum-product message-passing algorithm) and $\chi_{s_\mu}^{\mu}$ is the estimated posterior marginal on $s_\mu$, that we are interested in.

The AMP--BP algorithm reads:
\begin{multicols}{2}
\hrule
\vspace{2pt}
AMP--BP
\vspace{3pt}
\hrule height 0.7pt
\begin{algorithmic}
 \INPUT features $F_{\mu l}$, graph $G$, affinity matrix $c_{s,t}$, prior information $P_{s,\mu}$.
 \STATE Initialize $a_l^{(0)}=\epsilon_l$, $v_l^{(0)}=1$, $g_{o,\mu}^{(0)}=0$, $\chi_{s_\mu}^{\mu\to\nu, (0)}=\frac{1}{2}+s_\mu\epsilon^{\mu\to\nu}$, $\chi_{s_\mu}^{\mu\to\mu, (0)}=\frac{1}{2}$, $\chi_{s_\mu}^{\mu, (0)}=\frac{1}{2}$, $t=0$; where $\epsilon$s are zero-mean small random variables.
 \REPEAT
 \STATE AMP update of $\omega_\mu, V_\mu$ \\
\begin{align*}
& V^{(t+1)}\gets \frac{1}{M}\sum_lv_l^{(t)} \nonumber \\
&\omega_\mu^{(t+1)}\gets \sum_lF_{\mu l}a_l^{(t)} - V^{(t+1)}g_{o,\mu}^{(t)}
\end{align*}
 \STATE AMP update of $\psi^{\mu\to\mu}, g_{o,\mu}, \Lambda, \Gamma_l$\\
\begin{align*}
& \psi_{s_\mu}^{\mu\to\mu, (t+1)}\gets \int\frac{\mathrm dz P_0(s_\mu|z)}{\sqrt{2\pi V^{(t+1)}_\mu}}e^{-\frac{(z-\omega^{(t+1)}_\mu)^2}{2V^{(t+1)}_\mu}} \\
& g_{o,\mu}^{(t+1)}\gets g_o(\omega_\mu^{(t+1)}, \chi^{\mu\to\mu, (t)}, V^{(t+1)})
\end{align*}
\begin{align*}
& \Lambda^{(t+1)}\gets \frac{1}{M}\sum_\mu g_{o,\mu}^{2,(t+1)} \\
& \Gamma_l^{(t+1)}\gets \Lambda^{(t+1)}a_l^{(t)} + \sum_\mu F_{\mu l} g_{o,\mu}^{(t+1)}
\end{align*}
 \STATE AMP update of the estimated marginals $a_l, v_l$ \\
\begin{align*}
&a_l^{(t+1)}\gets f_a(\Lambda^{(t+1)}, \Gamma_l^{(t+1)})  \\
&v_l^{(t+1)}\gets f_v(\Lambda^{(t+1)}, \Gamma_l^{(t+1)})
\end{align*}
 \STATE BP update of the field $h$ 
\begin{align*}
&h_s^{(t+1)}\gets \frac{1}{N}\sum_\mu\sum_{s_\mu}c_{s,s_\mu}\chi_{s_\mu}^{\mu,(t)}
\end{align*}
 \STATE BP update of the messages $\chi^{\mu\to\nu}$ for $(\mu\nu)\in G$ and of the marginals $\chi^{\mu}$ 
\begin{align*}
\chi_{s_\mu}^{\mu\to\nu, (t+1)}\gets &\frac{P_{s,\mu}(s_\mu)}{Z^{\mu\to\nu}}e^{-h_{s_\mu}^{(t+1)}} \mkern-6mu \psi_{s_\mu}^{\mu\to\mu, (t+1)} \nonumber \\
 &\; \prod_{\eta\in\partial\mu\backslash \nu} \sum_{s_\eta}c_{s_\eta,s_\mu}\chi_{s_\eta}^{\eta\to\mu, (t)} \nonumber \\
\chi_{s_\mu}^{\mu, (t+1)}\gets &\frac{P_{s,\mu}(s_\mu)}{Z^{\mu}}e^{-h_{s_\mu}^{(t+1)}} \mkern-6mu \psi_{s_\mu}^{\mu\to\mu, (t+1)} \nonumber \\
 &\quad \prod_{\eta\in\partial\mu}\sum_{s_\eta}c_{s_\eta,s_\mu}\chi_{s_\eta}^{\eta\to\mu, (t)} \nonumber
\end{align*}
 \STATE BP update of the SBM-to-GLM messages $\chi^{\mu\to\mu}$ 
\begin{align*}
\chi_{s_\mu}^{\mu\to\mu, (t+1)}\gets &\frac{P_{s,\mu}(s_\mu)}{Z^{\mu\to\mu}}e^{-h_{s_\mu}^{(t+1)}} \nonumber \\
 &\quad \prod_{\eta\in\partial\mu}\sum_{s_\eta}c_{s_\eta,s_\mu}\chi_{s_\eta}^{\eta\to\mu, (t)}
\end{align*}
 \STATE $t\gets t+1$
 \UNTIL{convergence of $a_l, v_l, \chi^\mu$}
 \OUTPUT estimated mean $a_l$ and variance $v_l$ of $w_l$ and marginal distribution $\chi^\mu$ of $s_\mu$.
\end{algorithmic}
\hrule
\end{multicols}

We provide an implementation of AMP--BP in the supplementary material. It is also available from our repository.\footnote{\href{https://gitlab.epfl.ch/spoc-idephics/glm-sbm}{gitlab.epfl.ch/spoc-idephics/glm-sbm}} 

We draw attention to the output function $g_o$ that covers the difference between AMP for GLM--SBM and AMP for GLM alone. In AMP for GLM alone $g_o$ depends on the observed labels while here we use their estimated marginals. On the other side, the difference between BP for GLM--SBM and BP for SBM alone are the messages $\psi^{\mu\to\mu}$ in BP update. $\psi^{\mu\to\mu}$ can be interpreted as the conditional probability of $s_\mu$ given $w$ without SBM.

\paragraph*{Estimators.}
The Bayes-optimal estimators of $s$ and $w$ are obtained according to eqs.~\eqref{eq:estimateurMMO} and \eqref{eq:estimateurMMSE}. Expressed using the AMP--BP messages they become
\begin{equation}
\hat s^{\rm AMP–BP}_\mu=\mathrm{sign}(2\chi_+^\mu-1) \quad,\quad \hat w^{\rm AMP–BP}_l=a_l
\end{equation}
where $\chi_+^\mu$ is the estimated marginal probability of the event $s_\mu=+1$ and $a_l$ is the estimated mean of $w_l$.

\paragraph*{Free entropy.}
We express also the free entropy $\phi$ in terms of the messages and variables of AMP--BP at the fixed point; it is called the Bethe free entropy $\phi_\mathrm{Bethe}$. The derivation from the factor graph is done in appendix \ref{derivationEntropie}. Up to a term that diverges with $N$ we obtain that the Bethe free entropy is
\begin{align}
\phi_\mathrm{Bethe} &= \phi_\mathrm{SBM}+\phi_\mathrm{GLM} \\
\phi_\mathrm{SBM} &= \frac{1}{N}\sum_\mu\log \sum_{s_\mu}P_{s,\mu}(s_\mu)e^{-h_{s_\mu}} \prod_{\eta\in\partial\mu} \sum_{s_\eta}c_{s_\eta,s_\mu}\chi_{s_\eta}^{\eta \to\mu} \nonumber \\
  & \quad{}-\frac{1}{N}\sum_{(\mu\nu)\in G}\log\sum_{s_\mu,s_\nu}c_{s_\mu,s_\nu}\chi_{s_\mu}^{\mu\to\nu}\chi_{s_\nu}^{\nu\to\mu} + \frac{c}{2} \\
\phi_\mathrm{GLM} &= \frac{1}{N}\sum_\mu \log\int\frac{\mathrm dz_\mu}{\sqrt{2\pi V_\mu}}\,\sum_{s_\mu}\chi_{s_\mu}^{\mu\to\mu}P_o(s_\mu|z_\mu)e^{-(z_\mu-\omega_\mu)^2/2V_\mu} \nonumber \\
 &\quad{}+ \frac{1}{N}\sum_l\log\int\mathrm dw_l P_w(w_l) e^{-\Lambda_lw_l^2/2+\Gamma_lw_l} \nonumber \\
 &\quad{}+ \frac{1}{N}\left(\sum_l\frac{\Lambda_l}{2}(a_l^2+v_l)-\Gamma_la_l+\sum_\mu\frac{(\omega_\mu-\sum_lF_{\mu l}a_l)^2}{2V_\mu}\right)
\end{align}

If the AMP--BP has more than one fixed point then the free entropy serves to select the fixed point of AMP--BP that corresponds to Bayes-optimal performance. It is the one with the largest free entropy that should be selected. 

We compare later the free entropy of the fixed point of AMP--BP to the free entropy of the fully informative point where $q_S=q_W=1$. We write it $\phi_\mathrm{info}$. At this point the messages are delta functions of the ground truth; we can derive $\phi_\mathrm{info}$ directly from the factor graph and it reads
\begin{equation}
\phi_\mathrm{info} = \frac{1}{\alpha}\mathbb E_{P_w}\log P_w + \frac{1}{N}\sum_{(\mu\nu)\in G}\log c_{s_\mu,s_\nu} - \frac{c}{2} -(1-\rho)\log 2
\end{equation}

\subsection{Asymptotic optimality conjecture.}
We conjecture that AMP--BP gives the Bayes-optimal estimator for GLM--SBM in the following sense.

We define the two possible initializations: (a) random initialization, where we initialize the messages randomly according to their prior distribution, adding no information, as described in the algorithm above; and (b) informed initialization, where we initialize the estimators to delta functions of the true values of $s$ and $w$.

We consider the fixed point of AMP--BP that has the largest Bethe free entropy $\phi_\mathrm{Bethe}$. We argue that it suffices to check the random and informed initializations to find all the relevant fixed points.

We conjecture that, asymptotically exactly, the AMP--BP fixed point that has the largest $\phi_\mathrm{Bethe}$ provides the Bayes-optimal estimators for the GLM--SBM model. Its overlap $q_S$ is asymptotically equal to the Bayes-optimal MMO overlap and $\phi_\mathrm{Bethe}$ is equal to $\phi$, with high probability as $N\to +\infty$. This is aligned with the same conjecture for BP and the standard SBM from \cite{decelle2011inference} and the proofs of this property for the AMP algorithm and the pure GLM model in \cite{barbier2019optimal}.

\section{Bayes-optimal estimation with AMP--BP and phase transitions}
\subsection{Gaussian prior, 2nd order transition to partial recovery}
In this subsection we consider the GLM prior $P_w$ to be a standard Gaussian. The GLM then produces binary labels, the group memberships, with the same probability of being in each of the groups. 

We conjecture that for this prior the fixed point of AMP--BP reached from random initialization always corresponds to the Bayes-optimal estimation and no computationally hard phase is present. We observe that the algorithm converges to the same fixed point for the two possible initializations. In Fig. \ref{oVsN} in the appendix \ref{sec:figuresSuppl} we illustrate that the system size we use is close enough to the thermodynamical limit in the sense that the change in the curves is small when the size is changed.

The accuracy AMP--BP achieves is depicted in Figs.~\ref{oVsAlpha} (unsupervised case) and \ref{oVsAlphaSS} (semi-supervised case, in appendix \ref{sec:figuresSuppl}). We observe that the larger the snr $\lambda$ or the aspect ratio $\alpha$ the better the recovery. The recovery is eased when community memberships are explained by a few features i.e. when $\alpha$ is large. 
\begin{figure}[ht]
 \centering
 \includegraphics[width=\linewidth]{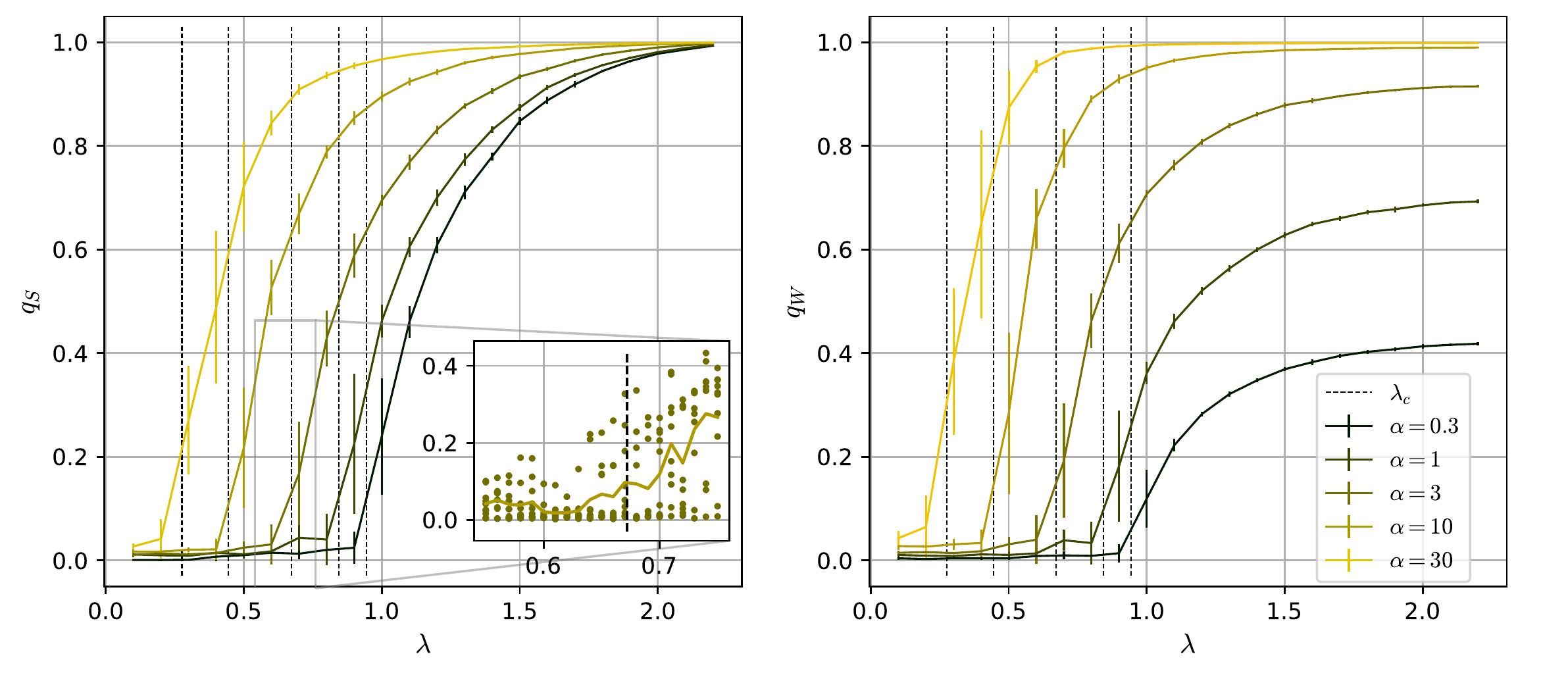}
 \caption{\label{oVsAlpha} \emph{Left and right:} overlaps $q_S$ (group membership estimation) and $q_W$ (the GLM latent vector estimation) of the fixed point of AMP--BP, vs $\lambda$ for a range of compression ratios $\alpha$. Vertical dashed lines: theoretical thresholds $\lambda_c$ to partial recovery, eq.~\eqref{eq:lambda_c}. $N=10^4$, $c=5$, $P_w$ Gaussian. We run ten experiments per point. \emph{Inset:} we plot the ten data points and their mean.}
\end{figure}

In the unsupervised case, we observe a phase transition from a non-informative fixed point $q_S=q_W=0$ to an informative fixed point $q_S>0, q_W>0$. The transition is located at a particular critical threshold $\lambda_c$. This transition is well known for standard SBM, which is recovered here in the $\alpha \to 0$ limit, where $\lambda_c=1$ for $q=2$. The transition is of 2nd order; this means that the overlaps vary continuously with respect to $\lambda$. In the semi-supervised case the 2nd order transition disappears.

\paragraph*{Linearization, spectral algorithm.} $\lambda_c$ can be computed by a linear stability analysis of the non-informative fixed point of AMP--BP: at a given $\lambda$, if the algorithm is not stable it will move away from the non-informative fixed point to the informative fixed point. The linearization of the algorithm is done in appendix \ref{linearisation}. We obtain the following update equation:
\begin{equation}
x^{\mu\to\nu,(t+1)} = \frac{\lambda}{\sqrt c}\left(\sum_{\eta\in\partial\mu\backslash\nu}x^{\eta\to\mu,(t)} + \frac{2}{\pi}\sum_{\eta}(FF^T-I_N)_{\mu,\eta}\sum_{\rho\in\partial\eta}x^{\rho\to\eta,(t-1)}\right) \label{spectralAlgorithm}
\end{equation}
where the $x$s are real random variables and $(FF^T)_{\mu\nu}=\sum_l F_{\mu l}F_{\nu l}$. Taking the variance of this equation and averaging over the realizations of the graph we obtain the stability criterion
\begin{equation}
\label{eq:lambda_c}
1 = \lambda_c^2\left(1+\frac{4\alpha}{\pi^2}\right)\, . 
\end{equation}

Eq. \eqref{spectralAlgorithm} can be interpreted as a spectral algorithm; we apply iteratively a linear operator to the variables
\begin{equation}
x = \frac{\lambda}{\sqrt c}\left(B+G\tilde B\right)x
\end{equation}
where
\begin{equation}
B_{\mu\to\nu,\rho\to\eta} = \delta_{\mu=\eta}(1-\delta_{\rho=\nu}) \quad,\quad G_{\mu\to\nu,\eta} = \frac{2}{\pi}(FF^T-I_N)_{\mu,\eta} \quad,\quad \tilde B_{\eta,\mu\to\rho} = \delta_{\rho=\eta}
\end{equation}
$B$ is the non-backtracking matrix \cite{SBM13spectral}. Such a spectral algorithm will share the phase transition at snr given by eq. \eqref{spectralAlgorithm}. The study of the resulting overlap is also of interest, but we do not consider it in the present article.

\subsection{Binary prior, 1st order transition to exact recovery}
\label{partBinaryPrior}
In this subsection the GLM prior is considered to be $P_w=(\delta_{w=1}+\delta_{w=-1})/2$ Rademacher. This still produces two groups with unbiased sizes. 

The fixed point AMP--BP achieves from a random initialization is depicted on Figs.~\ref{oVsAlphaBinaire} and \ref{oVsAlphaBinaireSS} (in appendix \ref{sec:figuresSuppl}). We observe it admits the same transition to partial recovery at $\lambda_c$, eq.~\eqref{eq:lambda_c}, as the Gaussian prior does; this is also predicted by the linearization of the previous part.
\begin{figure}[ht]
 \centering
 \includegraphics[width=\linewidth]{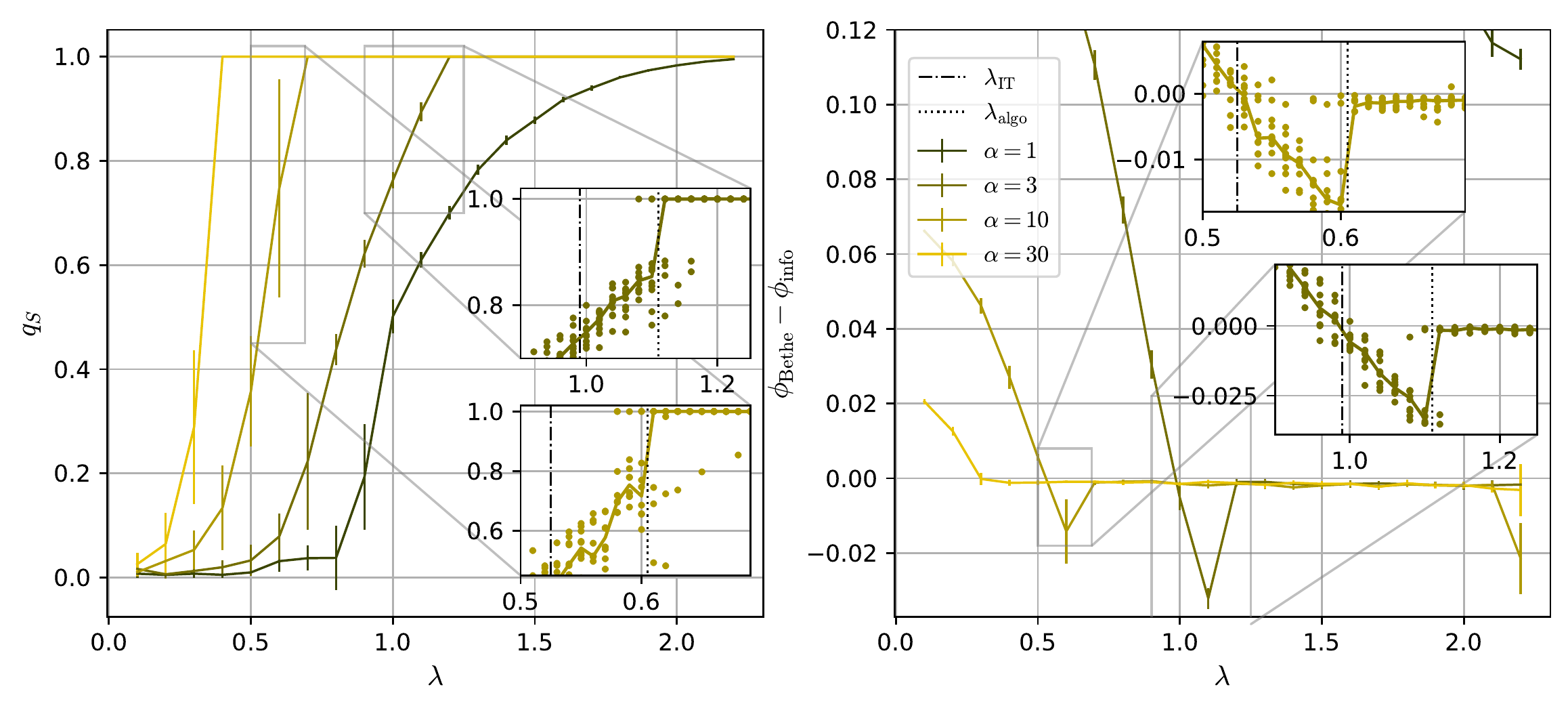}
 \caption{\label{oVsAlphaBinaire} \emph{Left and right:} overlap $q_S$ and free entropy $\phi_\mathrm{Bethe}-\phi_\mathrm{info}$ of the fixed point of AMP--BP, vs $\lambda$ for several compression ratios $\alpha$. $N=10^4$, $c=5$, $P_w$ Rademacher, $\rho=0$. We run ten experiments per point; the median is plotted and the error bars are the difference between the 0.85th and 0.15th quantiles. \emph{Insets:} we plot the median and the ten data points. We use damping for AMP--BP: we interpolate taking 1/4 of the values at $t+1$ and 3/4 of the values at $t$.
 }
\end{figure}

For values of $\alpha>\alpha_\mathrm{algo}$ (that we determine below), we observe another transition; it is discontinuous, from partial recovery $q_S>0, q_W>0$ to exact recovery $q_S=q_W=1$. There is a value $\lambda_\mathrm{algo}$ such that for $\lambda>\lambda_\mathrm{algo}$ randomly initialized AMP--BP recovers the group memberships exactly for all nodes. The overlap $q_S$ does not vary continuously at $\lambda_\mathrm{algo}$; over the many independent trials we observe that there is an interval of overlaps below 1 that cannot be reached by AMP--BP for any $\lambda$.  

Discontinuous thresholds are related to the existence of several fixed points of AMP--BP and to first-order phase transitions. A 1st order phase transition is located by comparing the free entropies $\phi_\mathrm{Bethe}$ of the various fixed points. We notice that next to the AMP--BP fixed point that is reached from a random initialization, the exact recovery point is a fixed point at all values of $\lambda$ and $\alpha$ (still considering $P_w$ binary). In the region of  $\lambda$ and $\alpha$  where these two fixed points differ we need to compare their free entropies. The fixed point with larger free entropy describes the Bayes-optimal performance that can in general be better than the one of AMP--BP. The difference between the free entropies of the fixed point reached by AMP--BP from random initialization and the informative fixed point is depicted on the rhs of Fig.~\ref{oVsAlphaBinaire}. We see that for $\lambda< \lambda_{\rm IT}$ the fixed point reached from random initialization has larger free entropy $\phi_\mathrm{Bethe} > \phi_\mathrm{info}$ and hence describes the optimal performance. In the region $\lambda_{\rm IT}< \lambda < \lambda_{\rm algo}$ the informative fixed point has larger free entropy $\phi_\mathrm{info}>\phi_\mathrm{Bethe}$, but randomly initialized AMP--BP does not reach it. This is an algorithmically hard phase where exact recovery is statistically possible, but the AMP--BP algorithm is sub-optimal. At the same time the AMP--BP algorithm is conjectured optimal among efficient algorithms \cite{gamarnik2022disordered} and thus the hardness of this phase is believed to be intrinsic. For $\lambda > \lambda_{\rm algo}$ we only find the exact recovery fixed point.

An exact recovery for the standard SBM is only achievable for graphs of average degrees $c$ diverging logarithmically with the size of the system \cite{abbe2015exact}, where the logarithm comes from a type of coupon collector problem. The existence of an exact recovery phase in graphs of constant degrees is novel as far as we know. It nicely illustrates the power of the GLM prior that is able to induce it. It is well known that a 1st order phase transition appears for GLM alone with binary weights and known labels \cite{gyorgyi1990first,sompolinsky1990learning,barbier2019optimal}. We note, however, that in the GLM--SBM the labels are not observed directly but via the graph. It is thus not a priori clear that an exact recovery phase can appear. Without our analysis its existence would not be easy to anticipate. 

Let us finally derive the values $\alpha_\mathrm{algo}$ above which the exact recovery phase exists. We consider the limit $\lambda=\sqrt c$; then the graph $G$ consists of two disconnected components, one for each community; and AMP--BP performs as AMP for GLM alone, up to a global sign. We take into account the proportion $e^{-c}$ of nodes that are isolated and do not bring information. We obtain that
\begin{equation}
\alpha_\mathrm{algo}=\alpha_\mathrm{algo,\,perceptron}\left(1-e^{-c}\right)^{-1}
\end{equation}
where $\alpha_\mathrm{algo,\,perceptron}\approx 1.493$ is the algorithmic critical compression ratio of the binary perceptron \cite{barbier2019optimal}. Similarly the $\lambda_{\rm IT}$ will exist above 
\begin{equation}
\alpha_\mathrm{IT}=\alpha_\mathrm{IT,\,perceptron}\left(1-e^{-c}\right)^{-1}
\end{equation}
where $\alpha_\mathrm{algo,\,perceptron}\approx 1.249$ is the information-theoretic critical threshold of the binary perceptron \cite{gyorgyi1990first,sompolinsky1990learning,barbier2019optimal}.

The 1st order phase transition $\lambda_{\rm IT}$ and its spinodal $\lambda_{\rm algo}$ are still present in the semi-supervised case $\rho>0$, for small values of $\rho$, see Fig.~\ref{oVsAlphaBinaireSS} in appendix~\ref{sec:figuresSuppl}, contrary to the 2nd order phase transition to partial recovery that vanishes in the semi-supervised case. Moreover, for $\rho>\alpha_\mathrm{algo,perceptron}/\alpha$, perfect recovery is achieved at any $\lambda$, because one has enough train labels to infer $w$.

\section{Analysis in the dense limit} 
\label{sec:limiteDense}
As in \cite{decelle2011inference,SBM11} for the sparse SBM, the analysis of the AMP--BP is based on the numerical investigation of the fixed points and their free entropies on systems large enough that the behavior is representative of the large-size limit. This is also at the basis of the mathematical difficulty to establish this prescription rigorously. At the same time, a dense version of the SBM has been proposed and studied fully rigorously in \cite{lesieur2017constrained,miolane2017fundamental}. This rigorous analysis has then been extended to include the GLM prior in \cite{aPrioriGen19} that studies a generic instance of low-rank matrix factorization problem with a generative prior. We hence study the phenomenology of the AMP--BP algorithm in the limit of large degree $c$, where it becomes a special case of the framework developed in \cite{aPrioriGen19}. 

The dense limit is defined by taking $p_i, p_o = \mathcal O(1)$ and $p_i-p_o = \mathcal O(1/\sqrt N)$. SBM is then a low-rank matrix factorization problem. It is parameterized by its signal-to-noise ratio $\Delta_I$ (which is defined as the inverse variance of an equivalent additive Gaussian channel). We need $\Delta_I$ as a function of the parameters of the SBM, that is to say to equalize their signal-to-noise ratios. We compute the Fisher information of the channel $P_{\rm SBM}(A_{\mu\nu}=1|x_{\mu\nu})=\frac{1}{N}c_o+\frac{1}{\sqrt N}(c_i-c_o)x_{\mu\nu}$, where $x_{\mu\nu}=\frac{\delta_{s_\mu=s_\nu}}{\sqrt N}$ is taken to zero. The mapping is then
\begin{equation}
\label{snr}
\Delta_I=\frac{1}{4}\frac{N(p_i-p_o)^2}{p_o(1-p_o)} = \frac{c\lambda^2}{c-\sqrt c\lambda}+\mathcal O\left(\frac{c}{N}\right)
\end{equation}
where $p_i=c_i/N$ and $p_o=c_o/N$. It is of order one in both sparse case and dense case. Also, we add the factor $1/4$ to obtain a phase transition at $\Delta_I=1$ in the dense case when $\alpha=0$. In the following $\rho=0$.

Authors of \cite{aPrioriGen19} give the algorithm corresponding to the dense case of AMP--BP algorithm. We reproduce it in appendix \ref{limiteDense}. Its performances can be tracked by a few scalar equations that are named state evolution (SE) equations. For $P_w$ Rademacher they read:
\begin{align}
q_w^{t+1} &= \mathbb E_\xi\left[ Z_w\left(\sqrt{\hat q^t_w}\xi, \hat q^t_w\right) f_w\left(\sqrt{\hat q^t_w}\xi, \hat q^t_w\right)^2 \right] \label{eq:SEdébut}\\
\hat q_w^t &= \alpha\mathbb E_{\xi,\eta}\left[ Z_o\left(\sqrt{q^t_s\Delta_I}\xi, q^t_s\Delta_I, \sqrt{q^t_w}\eta, 1-q^t_w\right) f_o\left(\sqrt{q^t_s\Delta_I}\xi, q^t_s\Delta_I, \sqrt{q^t_w}\eta, 1-q^t_w\right)^2 \right] \\
q_s^{t+1} &= \mathbb E_{\xi,\eta}\left[ Z_o\left(\sqrt{q^t_s\Delta_I}\xi, q^t_s\Delta_I, \sqrt{q^t_w}\eta, 1-q^t_w\right) f_s\left(\sqrt{q^t_s\Delta_I}\xi, q^t_s\Delta_I, \sqrt{q^t_w}\eta, 1-q^t_w\right)^2 \right] \label{eq:SEfin}
\end{align}
where $q_s$ and $q_w$ are the s- and w-overlaps, $\Delta_I$ is the signal-to-noise ratio of the problem, $\xi$ and $\eta$ are standard Gaussians and
\begin{align}
Z_w(\gamma, \Lambda) &= e^{\Lambda/2}\cosh\gamma \quad,\quad f_w(\gamma, \Lambda) = \partial_\gamma\log Z_w \\
Z_o(B, A, \omega, V) &= e^{-A/2}(\cosh(B)+\sinh(B)\mathrm{erf}(\omega/\sqrt{2V})) \\
f_o(B, A, \omega, V) &= \partial_\omega\log Z_o \quad,\quad f_s(B, A, \omega, V) = \partial_B\log Z_o
\end{align}
\cite{aPrioriGen19} gives also the free entropy of the fixed point of the algorithm for the dense problem. It reads
\begin{equation}
\phi_\mathrm{Bethe, d}(q_s, q_w, \hat q_w) = -\frac{\Delta_I}{4}q_s^2-\frac{1}{2\alpha}\hat q_wq_w+\psi_o(\Delta_I q_s, q_w)+\frac{1}{\alpha}\psi_w(\hat q_w)
\end{equation}
where
\begin{align}
\psi_o(\Delta_I q_s, q_w) &= \mathbb E_{\xi, \eta}\, \mathrm{xlogx}\, Z_o\left(\sqrt{\Delta_Iq_s}\xi, \Delta_Iq_s, \sqrt{q_w}\eta, 1-q_w\right) \\
\psi_w(\hat q_w) &= \mathbb E_\xi\, \mathrm{xlogx}\, Z_w\left(\sqrt{\hat q_w}\xi, \hat q_w\right)
\end{align}
xlogx being the function $x\to x\log x$.

\begin{figure}[ht]
 \centering
 \includegraphics[width=\linewidth]{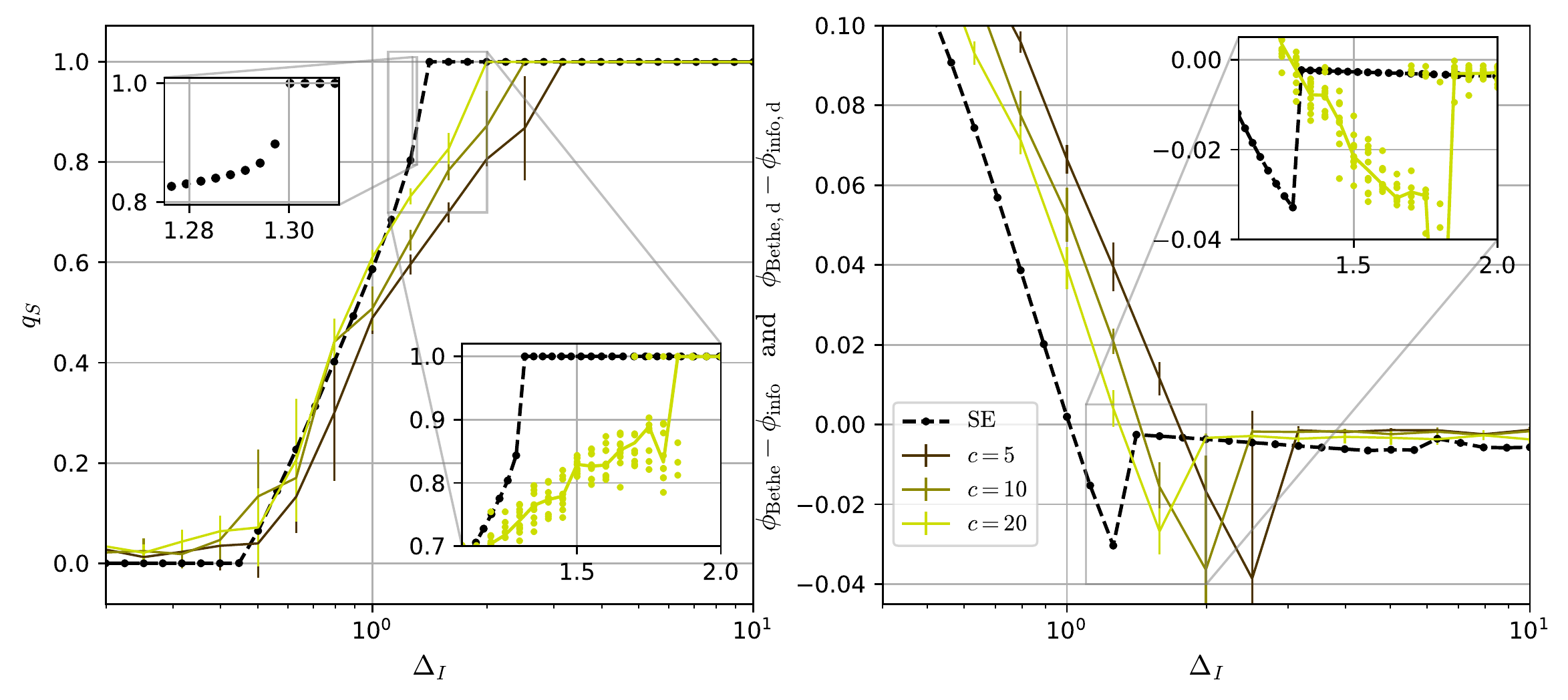}
 \caption{\label{oVsC} \emph{Left and right:} overlap $q_S$ and free entropies $\phi_\mathrm{Bethe}-\phi_\mathrm{info}$ and $\phi_\mathrm{Bethe,d}-\phi_\mathrm{info,d}$ of the fixed point of AMP--BP and of the SE equations of the dense limit, vs $\Delta_I$ for several average degrees $c$s. We generate instances of GLM--SBM according to the $\lambda$ obtained by inverting eq. \eqref{snr}. $N=10^4$, $\alpha=3$, $P_w$ binary. For AMP--BP we run ten experiments per point; for the SE equations one experiment. The median is plotted and the error bars are the difference between the 0.85th and 0.15th quantiles. \emph{Insets:} we plot the median and the ten data points. We use damping. For SE, the slight decrease of the free entropy at large $\Delta_I$ is due to numerical imprecision.}
\end{figure}

The convergence to the dense limit is quite fast; the large degree results are close to the observed results even for $c$ quite small. Numerically it appears that $c\approx 20$ is enough ($N=10^4$) to already observe quite small difference, see Fig.~\ref{oVsC}.

The fully informative fixed point is $(q_s, \hat q_w, q_w)=(1, +\infty, 1)$. Its free entropy is
\begin{equation}
\phi_\mathrm{info,d}=-\frac{\log 2}{\alpha}+\frac{\Delta_I}{4}
\end{equation}
The analysis of the system of SE equations is done in appendix \ref{limiteDense}; we summarize the four main points: (a) the fully informative fixed point is stable for all $\Delta_I$; (b) the width of its stability domain shrinks to zero when $\Delta_I$ tends to zero; (c) a general necessary condition to observe a fully informative fixed point is that $P_w$ does not admit everywhere a twice differentiable density; (d) the algorithmic critical compression ratio $\alpha_\mathrm{algo,d}$ is close to $\alpha_\mathrm{algo,preceptron}$.

We also obtain an approximation for the critical point $\lambda_c$ of the transition to partial recovery. \cite{aPrioriGen19} gives us that in the dense limit, the critical snr is
\begin{equation}
\Delta_{I,c}=\left(1+4\frac{\alpha}{\pi^2}\right)^{-1}
\end{equation}
The limit $c=\omega(1)$ large gives $\lambda_c=\left(1+{4}\alpha/{\pi^2}\right)^{-1/2} +\mathcal O(c/N)$, as predicted by the linearization.

\section{Comparison of performance with standard GNNs on GLM--SBM}
\label{comparaisonGLM}
GLM--SBM can be used as a benchmark for clustering or classification tasks on attributed graphs. We compare two simple baselines with AMP--BP. We show that GLM--SBM is simple to define yet challenging algorithmically, in particular in the case of binary prior close to the first order phase transition.

\paragraph*{An unsupervised baseline.} The algorithm is inspired by graph convolution networks; it performs binary clustering. We compare its performances to the optimal ones given by AMP--BP. Its performances are shown on Figs.~\ref{baselineBinary} left ($P_w$ binary) and \ref{baselineGaussian} left ($P_w$ Gaussian, in appendix \ref{sec:figuresSuppl}).

Data is generated according to the GLM--SBM. We stack the features $F_{\mu l}$ into vectors $F_\mu^{(0)}\in\mathbb R^M$ or a matrix $F\in\mathbb R^{N\times M}$. The observed graph $G$ is used for the convolution steps. 

We compute $n$ steps of graph convolution on the features; perform PCA on the transformed features and keep the largest component; threshold its left vector to obtain the membership of each node. Formally, we consider the features $F_\mu^{(0)}\in\mathbb R^M$; we apply $n$ times
\begin{equation}
F_\mu^{(t+1)}=F_\mu^{(t)}+a\sum_{\nu\in\partial\mu}F_\nu^{(t)}
\end{equation}
where $a$ is a scalar. We apply PCA on the new matrix $\hat F$ whose rows are $F_\mu^{(n)}$. Writing $u\in\mathbb R^N$ the left vector of its largest component, the estimator is $\hat s=\mathrm{sign}(u)$. We tune $n$ and $a$ empirically to optimize the recovery. We observe that roughly it depends on $n$ and $a$ only by their product $an$. Also, the optimal $a$ scales like $1/c$.
\begin{figure}[ht]
 \centering
 \includegraphics[width=0.5\linewidth]{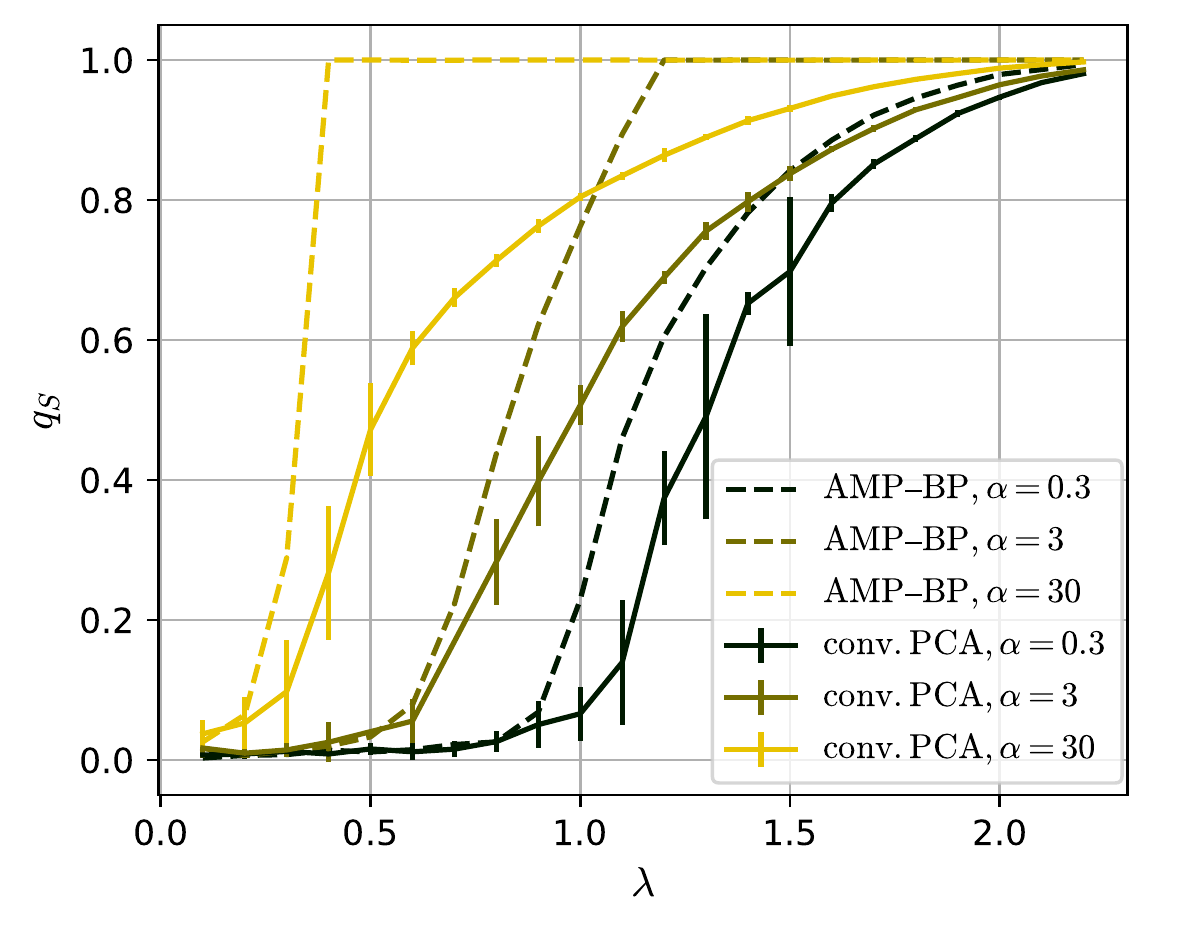}
 \includegraphics[width=0.48\linewidth]{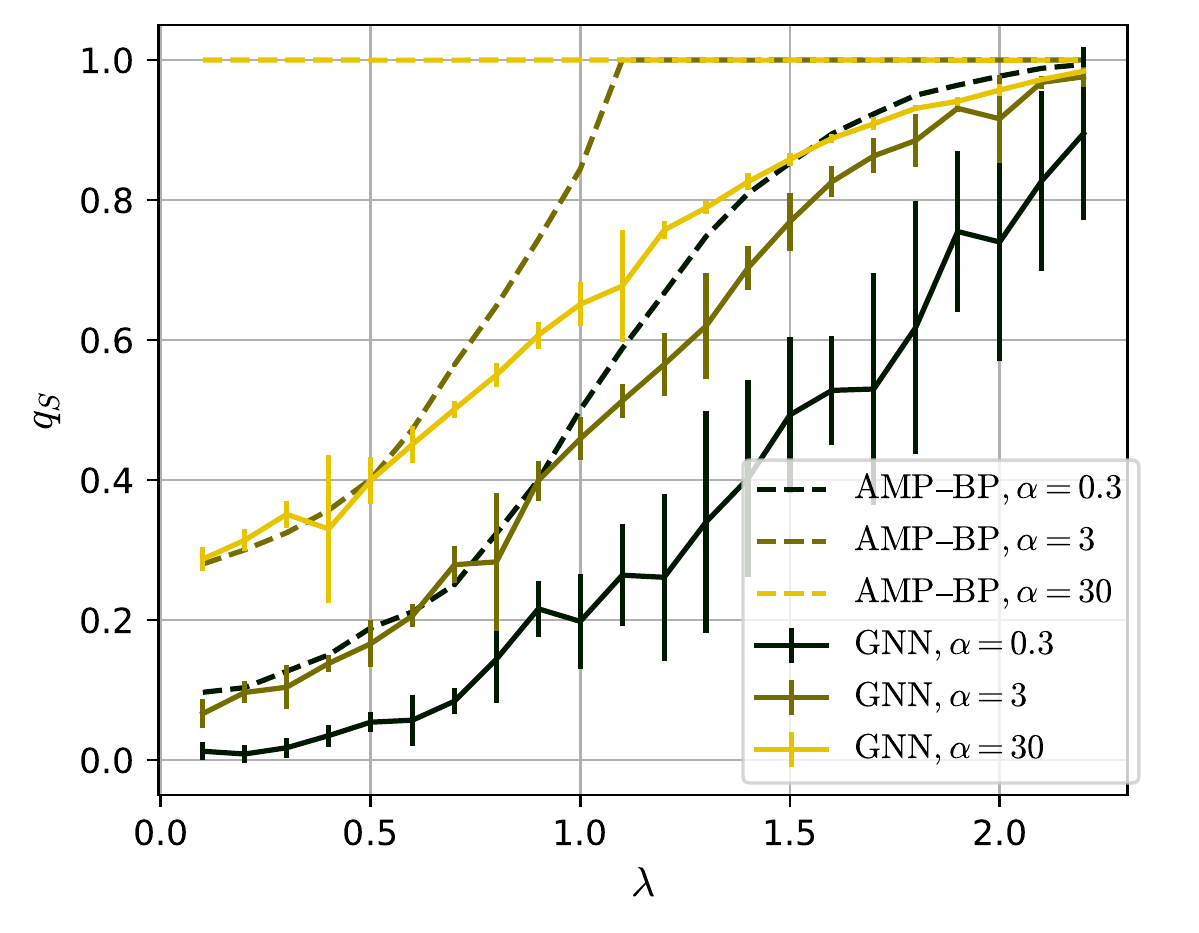}
 \caption{\label{baselineBinary} Overlap $q_S$ of the baseline algorithms, vs $\lambda$. We compare to the overlap obtained by AMP--BP. \emph{Left:} unsupervised; for the parameters of the graph convolution we choose $a=0.1$ and $n=4$. \emph{Right:} semi-supervised; for the hyperparameters of the GNN we choose $n=2$, $N_\mathrm{hidden}=20$, learning rate $3.10^{-4}$ and L2 penalty $10^{-3}$. The train set is $\rho=1/10^\mathrm{th}$ of the nodes. $N=10^4$, $c=5$, $P_w$ binary. We run ten experiments per point.}
\end{figure}

\paragraph*{A semi-supervised baseline.} The algorithm is a simple GNN, trained in a semi-supervised way for node classification.

Again data is generated according to the GLM--SBM, with $\rho=1/10$. We stack the features $F_{\mu l}$ into vectors $F_\mu^{(0)}\in\mathbb R^M$. We use the observed graph $G$ for the message-passing steps.

The GNN is made of a two-layer perceptron and a readout layer for the binary classification. It reads:
\begin{equation}
F_\mu^{(t+1)}=F_\mu^{(t)}+B\,\mathrm{relu}\left(A\sum_{\nu\in\partial\mu}F_\nu^{(t)}\right) \quad,\quad \hat s_\mu=w^TF_\mu^{(n)}
\end{equation}
where $A$ is $N_\mathrm{hidden}\times M$ learnable, $B$ is $M\times N_\mathrm{hidden}$ learnable, $w\in\mathbb R^M$ learnable and $n$ is the number of steps. We train it given the labels of the subset of nodes $\Xi$. We use gradient descent with logistic loss, momentum and L2 regularization. We do not fine-tune the hyperparameters. Its performances are shown on Figs.~\ref{baselineBinary} right ($P_w$ binary) and \ref{baselineGaussian} right ($P_w$ Gaussian, in appendix \ref{sec:figuresSuppl}).

We also performed experiments where the GNN is made of a single-layer perceptron (no relu), as \cite{GCNRepliques22} does on CSBM. The performances are similar to the multi-layer perceptron, but it requires much more parameters to be trained ($M^2$ vs $MN_\mathrm{hidden}$, and we take $N_\mathrm{hidden}=\mathcal O(1)$).

\paragraph*{Conclusion on the comparison.} As to the GLM--SBM dataset, Fig.~\ref{baselineBinary} illustrates that, both in the unsupervised and the semi-supervised settings, the baseline methods have a considerable gap to the optimal performances given by the AMP--BP algorithm. The GLM--SBM setting is hence suitable to develop GNN algorithms that are able to provide higher accuracy.

As to the AMP--BP algorithm, it is very scalable. It has a running time similar to the GNN-based approaches, around a few minutes per point on Figs. \ref{oVsAlpha} or \ref{oVsAlphaBinaire} (including the ten experiments). Its complexity is $\mathcal O(NM)$ in time and in memory. This is the smallest any algorithm can do, for reading the input. The number of steps needed for convergence does not depend on $N$.

\section{Conclusions}
We propose a model of attributed graphs. It is a sparse SBM where the nodes carry features that determine their community memberships. We solve it, in the sense that we derive an algorithm that is conjectured to perform optimally among polynomial algorithms.  We analyze a linearization of the algorithm and the dense limit of the model. The model, yet simple, exhibits a rich phenomenology with detectable and exact recovery phase transitions. It can be used as a challenging benchmark for graph-neural networks.

In the analysis of this paper we only considered two groups. For more than two groups, $q>2$, the analysis can also be done by writing an AMP--BP algorithm; just, the AMP-side would need to correspond to a single-layer network with multi-class output. The AMP for such a model has been written and studied in \cite{cornacchia2022learning} and one would have to merge it with the BP of \cite{decelle2011inference}. Another generalization that would be possible to analyze is when the attributes $F$ are drawn from a Gaussian with a generic covariance. This can be done along the lines of \cite{loureiro2021learning}. On the other hand considering as a prior the multi-layer neural network \eqref{eq:multi} with learned weights $W$ would be more challenging; a corresponding AMP algorithm that would provide an asymptotically exact solution is not known. 

A future direction of work could also be to theoretically analyze the learning of GLM--SBM by a GNN, i.e. to give insights on the generalization performance of the neural network of part \ref{comparaisonGLM}; as \cite{GCNRepliques22} does for a perceptron-based graph convolution network on CSBM. This would be interesting because few theoretical works address the generalization ability of GNNs.

\acks{We acknowledge funding from the ERC under the European Union’s Horizon 2020 Research and Innovation Program Grant Agreement 714608-SMiLe.}

\appendix

\section{Derivation of the algorithm}
\label{derivationAlgo}
We write belief propagation for this problem. We start with the factor graph. It contains six different messages:\\
\centerline{
\xymatrix{
 *+[o][F-]{w_l} \ar@{-}[r]\ar@{-}[dr] \ar@{}[r]^(.25){}="a"^(.75){}="b" \ar@<3pt>"a";"b"^{\chi_w^{l\to\mu}} & *+[F]{ } \ar@{-}[r] \ar@{}[r]^(.25){}="a"^(.75){}="b" \ar@<3pt>"a";"b"^{\psi_s^{\mu\to\mu}} & *+[o][F-]{s_\mu} & {}\save[]+<0cm,-0.65cm>*+[F]{ }  \ar@{-}[l] \ar@{}[l]^(.25){}="a"^(.75){}="b" \ar@<3pt>"b";"a"^{\chi_s^{\mu\to\nu}} \ar@{-}[ld] \ar@{}[ld]^(.25){}="a"^(.75){}="b" \ar@<3pt>"a";"b"^{\psi_s^{\mu\to\nu}} \restore \\
 *+[o][F-]{w_m} \ar@{-}[r]\ar@{-}[ur] \ar@{}[r]^(.25){}="a"^(.75){}="b" \ar@<3pt>"b";"a"^{\psi_w^{\nu\to m}} & *+[F]{ } \ar@{-}[r] \ar@{}[r]^(.25){}="a"^(.75){}="b" \ar@<3pt>"b";"a"^{\chi_s^{\nu\to\nu}} & *+[o][F-]{s_\nu} &
}
}
These messages satisfy these equations:
\begin{align}
\chi_{w_l}^{l\to\mu} &\propto P_w(w_l)\prod_{\nu\neq\mu} \psi_{w_l}^{\nu \to l}\\
\psi_{w_l}^{\nu\to l} &\propto \sum_{s_\nu}\chi_{s_\nu}^{\nu \to\nu} \int\prod_{m\neq l}\left(\mathrm dw_m \chi_{w_m}^{m\to\nu}\right) P_0(s_\nu | F_\nu.w)\\
\psi_{s_\nu}^{\nu\to\nu} &\propto \int \prod_m\left(\mathrm dw_m \chi_{w_m}^{m \to\nu}\right) P_0(s_\nu | F_\nu.w)\\
\chi_{s_\mu}^{\mu\to\mu} &\propto P_{s,\mu}(s_\mu)\prod_{\nu\neq\mu} \psi_{s_\mu}^{\nu\to\mu}\\
\chi_{s_\mu}^{\mu\to\nu} &\propto P_{s,\mu}(s_\mu)\psi_{s_\mu}^{\mu\to\mu}\prod_{\eta\neq\mu,\nu} \psi_{s_\mu}^{\eta\to\mu}\\
\psi_{s_\nu}^{\mu\to\nu} &\propto \sum_{s_\mu}\chi_{s_\mu}^{\mu\to\nu} P_{\rm SBM}(A_{\mu\nu} | s_\mu,s_\nu)
\end{align}

We can plug $\psi$ messages into the $\chi$ to obtain, for the GLM part:
\begin{equation}
\chi_{w_l}^{l\to\mu} = \frac{P_w(w_l)}{Z^{l\to\mu}}\prod_{\nu\neq\mu} \left[ \sum_{s_\nu}\chi_{s_\nu}^{\nu\to\nu} \int \prod_{m\neq l}\left(\mathrm dw_m \chi_{w_m}^{m \to\nu}\right) P_0(s_\nu | F_\nu.w) \right]\
\end{equation}
the marginals
\begin{equation}
\chi_{w_l}^{l} = \frac{P_w(w_l)}{Z^{l}}\prod_{\nu} \left[ \sum_{s_\nu}\chi_{s_\nu}^{\nu\to\nu} \int \prod_{m\neq l}\left(\mathrm dw_m \chi_{w_m}^{m \to\nu}\right) P_0(s_\nu | F_\nu.w) \right]\
\end{equation}
where the $Z$s are normalization factors; and for the SBM part:
\begin{equation}
\chi_{s_\mu}^{\mu\to\nu} \propto P_{s,\mu}(s_\mu)\psi_{s_\mu}^{\mu\to\mu} \prod_{\eta\neq\mu,\nu} \sum_{s_\eta}\chi_{s_\eta}^{\eta\to\mu} P_{\rm SBM}(A_{\mu\eta} | s_\mu,s_\eta)
\end{equation}
the marginals
\begin{equation}
\chi_{s_\mu}^{\mu} \propto P_{s,\mu}(s_\mu)\psi_{s_\mu}^{\mu\to\mu} \prod_{\eta\neq\mu} \sum_{s_\eta}\chi_{s_\eta}^{\eta\to\mu} P_{\rm SBM}(A_{\mu\eta} | s_\mu,s_\eta)
\end{equation}
and
\begin{equation}
\chi_{s_\mu}^{\mu\to\mu} \propto P_{s,\mu}(s_\mu)\prod_{\nu\neq\mu} \sum_{s_\nu}\chi_{s_\nu}^{\nu\to\mu} P_{\rm SBM}(A_{\mu\nu} | s_\mu,s_\nu)
\end{equation}

\subsection{SBM}
We can apply the standard simplifications for sparse SBM \cite{SBM11}, \cite{inferences15}. We consider only messages on $G$. This gives
\begin{align}
\chi_{s_\mu}^{\mu\to\mu} &= \frac{1}{Z^{\mu\to\mu}}  P_{s,\mu}(s_\mu)e^{-h_{s_\mu}} \prod_{\eta\in\partial\mu} \sum_{s_\eta}c_{s_\eta,s_\mu}\chi_{s_\eta}^{\eta \to\mu} \\
\chi_{s_\mu}^{\mu\to\nu} &= \frac{1}{Z^{\mu\to\nu}} P_{s,\mu}(s_\mu)\psi_{s_\mu}^{\mu\to\mu} e^{-h_{s_\mu}} \prod_{\eta\in\partial\mu\backslash \nu}\sum_{s_\eta}c_{s_\eta,s_\mu}\chi_{s_\eta}^{\eta\to\mu}
\end{align}
and the marginals
\begin{equation}
\chi_{s_\mu}^{\mu} = \frac{1}{Z^{\mu}} P_{s,\mu}(s_\mu)\psi_{s_\mu}^{\mu\to\mu} e^{-h_{s_\mu}} \prod_{\eta\in\partial\mu}\sum_{s_\eta}c_{s_\eta,s_\mu}\chi_{s_\eta}^{\eta\to\mu}
\end{equation}
where $h_s=\frac{1}{N}\sum_\mu\sum_{s_\mu}c_{s,s_\mu}\chi_{s_\mu}^\mu$.

\subsection{GLM}
For the GLM, we follow closely \cite{inferences15}.
\subsubsection{r-BP}
We apply first the simplifications that lead to r-BP. We define and consider the inner part of the $\chi_{w_l}^{l\to\mu}$ message:
\begin{equation}
\tilde \psi_{w_l}^{\nu\to l} = \sum_{s_\nu}\chi_{s_\nu}^{\nu \to\nu} \int\prod_{m\neq l}\left(\mathrm dw_m \chi_{w_m}^{m\to\nu}\right) P_0(s_\nu | F_\nu.w)
\end{equation}
We set $z_\nu=F_{\nu l}w_l+\sum_{m\neq l}F_{\nu m}w_m$. By independence of the $w$ the partial sum behaves like a Gaussian with mean and variance
\begin{equation}
\omega_{\nu\to l}=\sum_{m\neq l}F_{\nu m}a_{m\to\nu} \quad,\quad V_{\nu\to l}=\sum_{m\neq l}F^2_{\nu m}v_{m\to\nu}
\end{equation}
with
\begin{equation}
a_{m\to\nu}=\int\mathrm dw_m\chi_{w_m}^{m\to\nu}w_m \quad v_{m\to\nu}=\int\mathrm dw_m\chi_{w_m}^{m\to\nu}w^2_m - a_{m\to\nu}^2
\end{equation}
We replace the integral over all $w$s by a Gaussian integral over $z_\nu$; we obtain
\begin{equation}
\tilde\psi_{w_l}^{\nu\to l} = \sum_{s_\nu}\chi_{s_\nu}^{\nu \to\nu} \int\frac{\mathrm dz_\nu}{\sqrt {2\pi V_{\nu\to l}}} e^{-(z_\nu-F_{\nu l}w_l-\omega_{\nu\to l})^2/2V_{\nu\to l}} P_0(s_\nu | z_\nu)
\end{equation}
We can simplify. $F_{\nu l}$ is small, we expand the exponential:
\begin{equation}
e^{-\frac{(z_\nu-F_{\nu l}w_l-\omega_{\nu\to l})^2}{2V_{\nu\to l}}} = e^{-\frac{(z_\nu-\omega_{\nu\to l})^2}{2V_{\nu\to l}}}\left(1-\frac{F_{\nu l}^2w_l^2}{2V_{\nu\to l}}+\frac{(z_\nu-\omega_{\nu\to l})F_{\nu l}w_l}{V_{\nu\to l}}+\frac{(z_\nu-\omega_{\nu\to l})^2F_{\nu l}^2w_l^2}{2V_{\nu\to l}^2}\right)
\end{equation}
We introduce the denoising function; its expression differs from the one of \cite{inferences15}:
\begin{equation}
g_o(\omega, \chi, V) = \frac{\int\mathrm dz\sum_s\chi_s P_0(s|z)(z-\omega)e^{-(z-\omega)^2/2V}}{V\int\mathrm dz\sum_s\chi_s P_0(s|z)e^{-(z-\omega)^2/2V}}
\end{equation}
So 
\begin{equation}
\tilde\psi_{w_l}^{\nu\to l} \propto \left(1-\frac{F_{\nu l}^2w_l^2}{2V_{\nu\to l}}+g_oF_{\nu l}w_l+\frac{1}{2}(\frac{1}{V_{\nu\to l}}+\partial_\omega g_o+g_o^2)F_{\nu l}^2w_l^2 \right)
\end{equation}
where we evaluate $g_o$ in $(\omega_{\nu\to l}, \chi^{\nu \to\nu}, V_{\nu\to l})$.
We exponentiate:
\begin{equation}
\tilde\psi_{w_l}^{\nu\to l} \propto e^{g_oF_{\nu l}w_l+\frac{1}{2}\partial_\omega g_oF_{\nu l}^2w_l^2}
\end{equation}
We take the product of the $\tilde\psi$ to obtain
\begin{equation}
\chi_{w_l}^{l\to\mu} \propto P_w(w_l) e^{-\Lambda_{l\to\mu}w_l^2/2+\Gamma_{l\to\mu}w_l}
\end{equation}
where
\begin{equation}
\Lambda_{l\to\mu}=-\sum_{\nu\neq\mu}\partial_\omega g_o(\omega_{\nu\to l}, \chi^{\nu \to\nu}, V_{\nu\to l}) F^2_{\nu l} \quad,\quad \Gamma_{l\to\mu}=\sum_{\nu\neq\mu}g_o(\omega_{\nu\to l}, \chi^{\nu \to\nu}, V_{\nu\to l}) F_{\nu l}
\end{equation}
We close the loop defining the input functions
\begin{equation}
f_a(\Lambda, \Gamma)=\frac{\int\mathrm dw\,P_w(w)we^{-\Lambda w^2/2+\Gamma w}}{\int\mathrm dwP_w(w)e^{-\Lambda w^2/2+\Gamma w}} \quad,\quad f_v(\Lambda, \Gamma)=\partial_\Gamma f_a(\Lambda, \Gamma)
\end{equation}
so
\begin{equation}
a_{l\to\mu}=f_a(\Gamma_{l\to\mu}, \Lambda_{l\to\mu}) \quad,\quad v_{l\to\mu}=f_v(\Gamma_{l\to\mu}, \Lambda_{l\to\mu})
\end{equation}
The mean and the variance of the marginals are estimated by
\begin{equation}
a_l=f_a(\Gamma_l, \Lambda_l) \quad,\quad v_l=f_v(\Gamma_l, \Lambda_l)
\end{equation}
where
\begin{equation}
\Lambda_{l}=-\sum_{\nu}\partial_\omega g_o(\omega_{\nu\to l}, \chi^{\nu \to\nu}, V_{\nu\to l}) F^2_{\nu l} \quad,\quad \Gamma_{l}=\sum_{\nu}g_o(\omega_{\nu\to l}, \chi^{\nu \to\nu}, V_{\nu\to l}) F_{\nu l}
\end{equation}
We obtain also the expression of the GLM-to-SBM message
\begin{equation}
\psi_{s_\mu}^{\mu\to\mu} = \frac{1}{\sqrt{2\pi V_\mu}} \int\mathrm dz P_0(s_\mu|z)e^{-(z-\omega_\mu)^2/2V_\mu}
\end{equation}
where
\begin{equation}
\omega_{\mu}=\sum_{m}F_{\mu m}a_{m\to\mu} \quad,\quad V_{\mu}=\sum_{m}F^2_{\mu m}v_{m\to\mu}
\end{equation}

\subsubsection{Time indices}
There are two possibilities for mixing the GLM part and the SBM part:\\
\centerline{
\xymatrix{
a,v^{(t)} \ar[r] & \omega,V^{(t+1)} \ar[r]\ar[d] & \Gamma,\Lambda^{(t+1)} \ar[r] & a,v^{(t+1)} \\
 & \psi^{(t+1)} \ar[dr] & g_o^{(t+1)} \ar[u] & & \\
 & \chi^{(t)} \ar[r] & \chi^{(t+1)} \ar@{=>}[u] & &
}
}
or\\
\centerline{
\xymatrix{
a,v^{(t)} \ar[r] & \omega,V^{(t+1)} \ar[r]\ar[d] & \Gamma,\Lambda^{(t+1)} \ar[r] & a,v^{(t+1)} \\
 & \psi^{(t+1)} \ar[dr] & g_o^{(t+1)} \ar[u] & & \\
 & \chi^{(t)} \ar[r]\ar@{=>}[ur] & \chi^{(t+1)} & &
}
}
We try both; we do not observe any numerical difference.

\subsubsection{AMP}
Then we go from r-BP to AMP. We remove the dependence of the messages on the target. We keep only the marginals. The derivation is given by \cite{inferences15}. We obtain that
\begin{align}
V^{(t+1)}_\mu &= \sum_lF_{\mu m}^2v_l^{(t)} \label{V-AMP-F2} \\
\omega_\mu^{(t+1)} &= \sum_lF_{\mu l}a_l^{(t)} - V^{(t+1)}_\mu g_{o,\mu}^{(t)} \\
g_{o,\mu}^{(t+1)} &= g_o(\omega_\mu^{(t+1)}, \chi^{\mu\to\mu, (t)}, V_\mu^{(t+1)}) \\
\Lambda^{(t+1)}_l &= -\sum_\mu F_{\mu l}^2\partial_\omega g_o(\omega_\mu^{(t+1)}, \chi^{\mu\to\mu, (t)}, V^{(t+1)}_\mu) \label{Sigma-AMP-F2} \\
\Gamma_l^{(t+1)} &= \Lambda^{(t+1)}_la_l^{(t)} + \sum_\mu F_{\mu l} g_{o,\mu}^{(t+1)}
\end{align}

\subsubsection{Further simplifications}
$F_{\mu m}^2$ self-averages. We can replace it by its average $1/M$ in eqs. \eqref{V-AMP-F2} and \eqref{Sigma-AMP-F2}. So $\Lambda$ and $V$ become scalars. Also, on average, $-\partial_\omega g_{o,\mu}=g_{o,\mu}^2$. We obtain the algorithm given in the main part.

\section{Free entropy}
\label{derivationEntropie}
We start with the factor graph. The Bethe free entropy is the sum of the free entropies of the nodes plus the factors minus the edges i.e.
\begin{align*}
N\phi_\mathrm{Bethe} &= \sum_\mu \phi^\mu+\sum_l \phi^l+\sum_{\mu<\nu}\phi^{\mu\nu}+\sum_\mu \phi^{\mu\mu} \\
 & \quad {}-\sum_{\mu\neq\nu}\phi^{\mu\to\nu}-\sum_\mu \phi^{\mu\to\mu}-\sum_{l,\mu}\phi^{l\to\mu}
\end{align*}
where
\begin{align}
\phi^\mu &= \log\sum_{s_\mu}P_{s,\mu}(s_\mu)\psi_{s_\mu}^{\mu\to\mu}\prod_{\nu\neq\mu}\psi_{s_\mu}^{\nu\to\mu} = \log Z^\mu\prod_{\nu\neq\mu}\frac{1}{Z^{\nu\to\mu}_\psi} \\
\phi^l &= \log\int\mathrm dw_l\,P_w(w_l)\prod_\mu\psi_{w_l}^{\mu\to l} = \log Z^l\prod_{\mu}\frac{1}{Z^{\mu\to l}_\psi} \\
\phi^{\mu\nu} &= \log\sum_{s_\mu,s_\nu}\chi_{s_\mu}^{\mu\to\nu}\chi_{s_\nu}^{\nu\to\mu}P(A_{\mu\nu} | s_\mu,s_\nu) \\
\phi^{\mu\mu} &= \log\sum_{s_\mu} \int\prod_{m}\left(\mathrm dw_m\chi_{w_m}^{m\to\mu}\right)\chi_{s_\mu}^{\mu\to\mu} P_0(s_\mu | F_\mu.w)\\
\phi^{\mu\to\nu} &= \log\sum_{s_\mu}\chi_{s_\mu}^{\mu\to\nu}\psi_{s_\mu}^{\nu\to\mu} = \log\frac{1}{Z_\psi^{\nu\to\mu}}+\phi^{\mu\nu} \\
\phi^{\mu\to\mu} &= \log\sum_{s_\mu}\chi_{s_\mu}^{\mu\to\mu}\psi_{s_\mu}^{\mu\to\mu} = \log\frac{1}{Z^{\mu\to\mu}}+\phi^{\mu} \\
\phi^{l\to\mu} &= \log\int\mathrm dw_l\chi_{w_l}^{l\to\mu}\,\psi_{w_l}^{\mu\to l} = \log\frac{1}{Z_\psi^{\mu\to l}}+\phi^{\mu\mu}
\end{align}
This simplifies to
\begin{align}
N\phi_\mathrm{Bethe} &= \sum_\mu\log Z^{\mu\to\mu}+\sum_l\log Z^l-\sum_{\mu<\nu}\phi^{\mu\nu}+(1-M)\sum_\mu \phi^{\mu\mu}
\end{align}
On the SBM side, we have \cite{inferences15}
\begin{align}
Z^{\mu\to\mu} &= \sum_{s_\mu}P_{s,\mu}(s_\mu)e^{-h_{s_\mu}} \prod_{\eta\in\partial\mu} \sum_{s_\eta}c_{s_\eta,s_\mu}\chi_{s_\eta}^{\eta \to\mu} \\
\sum_{\mu<\nu}\phi^{\mu\nu} &= \sum_{(\mu\nu)\in G}\log\sum_{s_\mu,s_\nu}c_{s_\mu,s_\nu}\chi_{s_\mu}^{\mu\to\nu}\chi_{s_\nu}^{\nu\to\mu} -N\frac{c}{2}
\end{align}
On the GLM side, we have
\begin{align}
\phi^{\mu\mu} &= \log\int\frac{\mathrm dz_\mu}{\sqrt{2\pi V_\mu}}\,\sum_{s_\mu}\chi_{s_\mu}^{\mu\to\mu}P_o(s_\mu|z_\mu)e^{-(z_\mu-\omega_\mu)^2/2V_\mu} \\
\log Z^l &= \sum_\mu \log\hat Z^{\mu\to l} + \log\int\mathrm dw_l P_w(w_l) e^{-\Lambda_lw_l^2/2+\Gamma_lw_l}
\end{align}
We compute $\log\hat Z^{\mu\to l}$ as a function of the target-free elements (we start using that $\omega_{\mu\to l}=\omega_\mu-F_{\mu l}a_{l\to\mu}$ and $V_{\mu\to l}=V_\mu-F_{\mu l}^2v_{l\to\mu}$ and expanding). This gives:
\begin{align}
\sum_{l,\mu}\log\hat Z^{\mu\to l} &= M\sum_\mu \phi^{\mu\mu}+\sum_l\frac{\Lambda_l}{2}(a_l^2+v_l)-\Gamma_la_l+\sum_\mu\frac{(\omega_\mu-\sum_lF_{\mu l}a_l)^2}{2V_\mu}
\end{align}
which is what \cite{freeEntropyGLM14} gives (taking $\Sigma_l=1/\Lambda_l$ and $R_l=\Gamma_l/\Lambda_l$). Finally, we obtain that the free entropy $\phi$ is exactly the sum of the free entropies of the two sub-problems:
\begin{align}
\phi_\mathrm{Bethe} &= \phi_\mathrm{SBM}+\phi_\mathrm{GLM} \\
\phi_\mathrm{SBM} &= \frac{1}{N}\sum_\mu\log \sum_{s_\mu}P_{s,\mu}(s_\mu)e^{-h_{s_\mu}} \prod_{\eta\in\partial\mu} \sum_{s_\eta}c_{s_\eta,s_\mu}\chi_{s_\eta}^{\eta \to\mu} \nonumber \\
  & \quad{}-\frac{1}{N}\sum_{(\mu\nu)\in G}\log\sum_{s_\mu,s_\nu}c_{s_\mu,s_\nu}\chi_{s_\mu}^{\mu\to\nu}\chi_{s_\nu}^{\nu\to\mu} + \frac{c}{2} \\
\phi_\mathrm{GLM} &= \frac{1}{N}\sum_\mu \log\int\frac{\mathrm dz_\mu}{\sqrt{2\pi V_\mu}}\,\sum_{s_\mu}\chi_{s_\mu}^{\mu\to\mu}P_o(s_\mu|z_\mu)e^{-(z_\mu-\omega_\mu)^2/2V_\mu} \nonumber \\
 &\quad{}+ \frac{1}{N}\sum_l\log\int\mathrm dw_l P_w(w_l) e^{-\Lambda_lw_l^2/2+\Gamma_lw_l} \nonumber \\
 &\quad{}+ \frac{1}{N}\left(\sum_l\frac{\Lambda_l}{2}(a_l^2+v_l)-\Gamma_la_l+\sum_\mu\frac{(\omega_\mu-\sum_lF_{\mu l}a_l)^2}{2V_\mu}\right)
\end{align}

\section{Linearization and partial recovery threshold}
\label{linearisation}
We take $P_{s,\mu}(s)=1/2$. The non-informative point $q_S=q_W=0$ is a fixed point of the AMP--BP algorithm. At this point, we have $\chi^{\mu\to\nu}=\frac{1}{2}$, $\chi^{\mu\to\mu}=\frac{1}{2}$, $\chi^{\mu\to\nu}=\frac{1}{2}$, $a_l=0$, $v_l=1$, $\omega_\mu=0$, $V=1$, $\psi^{\mu\to\mu}=\frac{1}{2}$, $g_{o,\mu}=0$, $\Lambda=0$ and $\Gamma_l=0$.

We linearize the equations of the algorithm around this point. We write $|_*$ the evaluation of functions in this point. We have
\begin{align}
\delta\chi^{\mu\to\nu,(t+1)} &= \sum_{\eta\in\partial\mu\backslash\nu}\frac{1}{2}\left(\frac{c_{.,.}}{c}-1\right)\delta\chi^{\eta\to\mu,(t)} + \partial_\omega\psi^{\mu\to\mu}|_*\delta\omega_\mu^{(t+1)} + \partial_V\psi^{\mu\to\mu}|_*\delta V^{(t+1)} \\
\delta\chi^{\mu\to\mu,(t+1)} &= \sum_{\eta\in\partial\mu}\frac{1}{2}\left(\frac{c_{.,.}}{c}-1\right)\delta\chi^{\eta\to\mu,(t)} \\
\delta a_l^{(t+1)} &= \partial_\Lambda f_a|_*\delta\Lambda^{(t+1)} + \partial_\Gamma f_a|_*\delta\Gamma_l^{(t+1)} \\
\delta v_l^{(t+1)} &= \partial_{\Lambda\Gamma} f_a|_*\delta\Lambda^{(t+1)} + \partial_{\Gamma\Gamma} f_a|_*\delta\Gamma_l^{(t+1)} \\
\delta g_{o,\mu}^{(t+1)} &= \partial_\omega g_o|_*\delta\omega_\mu^{(t+1)} + \nabla_\chi g_o|_*\delta\chi^{\mu\to\mu,(t+1)} + \partial_V g_o|_*\delta V^{(t+1)}
\end{align}
where we write $c_{.,.}$ for the affinity matrix and where we have used the standard linearization for SBM. We have also
\begin{align}
\delta\omega_\mu^{(t+1)} &= \sum_lF_{\mu l}\delta a_l^{(t)} - \delta V^{(t+1)}g_o|_* - V|_*\delta g_{o,\mu}^{(t)} \\
\delta V^{(t+1)} &= \frac{1}{M}\sum_l\delta v_l^{(t)} \\
\delta \Lambda^{(t+1)} &= \frac{2}{M}\sum_\mu g_o|_*\delta g_{o,\mu}^{(t+1)} \\
\delta \Gamma_l^{(t+1)} &= \delta\Lambda^{(t+1)}a_l|_* + \Lambda|_*\delta a_l^{(t)} + \sum_\mu F_{\mu l}\delta g_{o,\mu}^{(t+1)}
\end{align}
We simplify: $g_o|_*=0$, $\partial_\omega g_o|_*=0$, $\partial_V g_o|_*=0$, $\partial_V\psi|_*=0$ and $\partial_\Gamma f_a|_*=1$ (for $P_W$ both Gaussian or Rademacher). We compute that $\partial_\omega\psi|_*=\frac{1}{\sqrt{2\pi}}\left(\begin{smallmatrix} 1\\ -1\end{smallmatrix}\right)$ and $\nabla_\chi g_o|_*=\frac{2}{\sqrt{2\pi}}\left(\begin{smallmatrix} 1\\ -1\end{smallmatrix}\right)^T$. We assemble equations together:
\begin{align}
\delta\chi^{\mu\to\nu,(t+1)} &= \sum_{\eta\in\partial\mu\backslash\nu}\frac{1}{2}\left(\frac{c_{.,.}}{c}-1\right)\delta\chi^{\eta\to\mu,(t)} + \partial_\omega\psi|_*\left(\sum_lF_{\mu l}\delta a_l^{(t)} - \nabla_\chi g_o|_*\delta\chi^{\mu\to\mu,(t)}\right) \\
\delta\chi^{\mu\to\nu,(t+1)} &= \sum_{\eta\in\partial\mu\backslash\nu}\frac{1}{2}\left(\frac{c_{.,.}}{c}-1\right)\delta\chi^{\eta\to\mu,(t)} \nonumber\\
 &\quad{}+ \sum_{\eta,l}F_{\mu l}F_{\eta l}(\partial_\omega\psi|_*.\nabla_\chi g_o|_*)\delta\chi^{\eta\to\eta,(t)} - (\partial_\omega\psi|_*\nabla_\chi g_o|_*)\delta\chi^{\mu\to\mu,(t)} \\
\delta\chi^{\eta\to\eta,(t)} &= \sum_{\rho\in\partial\eta}\frac{1}{2}\left(\frac{c_{.,.}}{c}-1\right)\delta\chi^{\rho\to\eta,(t-1)}
\end{align}
The matrices $\frac{1}{2}\left(\frac{c_{.,.}}{c}-1\right)$ and $\partial_\omega\psi|_*\nabla_\chi g_o|_*$ share the same eigenvectors. They have one null eigenvalue and one positive: $\frac{c_i-c_o}{2c}=\frac{\lambda}{\sqrt c}$ and $\frac{2}{\pi}$. We project to obtain 
\begin{align}
x^{\mu\to\nu,(t+1)} &= \frac{\lambda}{\sqrt c}\left(\sum_{\eta\in\partial\mu\backslash\nu}x^{\eta\to\mu,(t)} + \frac{2}{\pi}\sum_{\eta}(FF^T)_{\mu,\eta}\sum_{\rho\in\partial\eta}x^{\rho\to\eta,(t-1)} - \frac{2}{\pi}\sum_{\eta\in\partial\mu}x^{\eta\to\mu,(t-1)}\right) \\
x^{\mu\to\nu,(t+1)} &= \frac{\lambda}{\sqrt c}\left(\sum_{\eta\in\partial\mu\backslash\nu}x^{\eta\to\mu,(t)} + \frac{2}{\pi}\sum_{\eta}(FF^T-I_N)_{\mu,\eta}\sum_{\rho\in\partial\eta}x^{\rho\to\eta,(t-1)}\right) \label{spectralAlgorithmBis}
\end{align}
where $(FF^T)_{\mu\nu}=\sum_l F_{\mu l}F_{\nu l}$.

We obtain the threshold $\lambda_c$ of partial recovery taking the variance of the expression \ref{spectralAlgorithmBis}, discarding the time indices. We use that $(FF^T-I_N)^2_{\mu,\nu}$ averages to $1/M$ if $\mu\neq\nu$ and to $\mathcal O(1/M)$ otherwise. We obtain:
\begin{align}
1 &= \lambda_c^2\left(1+\frac{4\alpha}{\pi^2}\right)
\end{align}

\section{Dense limit}
\label{limiteDense}
We consider the limit $c$ large. GLM--SBM is equivalent to a low-rank matrix factorization problem with a generative prior. It has been studied in \cite{aPrioriGen19}. We follow it closely.

We set $p_o=c_o/N$ and $p_i=c_i/N$ of order one and $\mu=\sqrt N(p_i-p_o)$ of order one. The effective inverse noise of the SBM is \cite{lesieur2017constrained}
\begin{equation}
\Delta_I=\frac{\mu^2}{p_o(1-p_o)}
\end{equation}

\subsection{Algorithm}
We reproduce here the algorithm given by \cite{aPrioriGen19}; we simplify it for a binary output channel and $P_w$ Rademacher; and we complete it with the semi-supervised case. In the dense limit BP can be approximated by AMP and our algorithm AMP--BP becomes AMP--AMP.

We set, as in part \ref{sec:limiteDense}:
\begin{align}
Z_w(\gamma, \Lambda) &= e^{\Lambda/2}\cosh\gamma \quad,\quad f_w(\gamma, \Lambda) = \partial_\gamma\log Z_w \\
Z_{o,\mu}(B, A, \omega, V) &=
\left \{
\begin{array}{r c l}
  e^{-A/2+s_\mu B}\frac{1}{2}(1+s_\mu\mathrm{erf}(\omega/\sqrt{2V})) & \mathrm{if}\;s_\mu\in s_\Xi \\
  e^{-A/2}(\cosh(B)+\sinh(B)\mathrm{erf}(\omega/\sqrt{2V})) & \mathrm{else}
\end{array}
\right .\\
f_{o,\mu}(B, A, \omega, V) &= \partial_\omega\log Z_{o,\mu} \quad,\quad f_{s,\mu}(B, A, \omega, V) = \partial_B\log Z_{o,\mu}
\end{align}
We define the input matrix
\begin{equation}
S_{\eta\nu}=\mu\frac{1}{1-p_o}\left(\frac{Y_{\eta\nu}}{p_o}-1\right)
\end{equation}
where $Y$ is the observed adjacency matrix; $Y_{\eta\nu}=1$ if there is an edge between $\eta$ and $\nu$, 0 otherwise.

In the following $a_l$ and $v_l$ are estimators of the mean and the variance of $w_l$; $\sigma_\mu$ and $\Sigma_\mu$ the mean and the variance of $s_\mu$.
\begin{multicols}{2}
\hrule
\vspace{3pt}
AMP--AMP
\vspace{3pt}
\hrule height 0.7pt
\begin{algorithmic}
 \INPUT features $F_{\mu l}$, input matrix $S_{\eta\nu}$, effective inverse noise $\Delta_I$, prior information $P_{s,\mu}$.
 \STATE Initialize $a_l^{(0)}=\epsilon_l$, $v_l^{(0)}=1$, $\sigma_\nu^{(0)}=\epsilon_\nu$, $\Sigma_\nu^{(0)}=1$, $g_{o,\mu}^{(0)}=0$, $t=0$; where $\epsilon$s are zero-mean small random variables.
 \REPEAT
 \STATE AMP update of $\omega_\mu, V_\mu$ \\
\begin{align*}
& V^{(t+1)}\gets \frac{1}{M}\sum_lv_l^{(t)} \\
& \omega_\mu^{(t+1)}\gets \sum_lF_{\mu l}a_l^{(t)} - V^{(t+1)}g_{o,\mu}^{(t)}
\end{align*}
 \STATE AMP update of $g_{o,\mu}, \Lambda, \Gamma_l$\\
\begin{align*}
g_{o,\mu}^{(t+1)}\gets f_{o,\mu}&(B_\mu^{(t+1)}, A^{(t+1)}, \\
&\quad \omega_\mu^{(t+1)}, V^{(t+1)})
\end{align*}
\begin{align*}
& \Lambda^{(t+1)}\gets \frac{1}{M}\sum_\mu g_{o,\mu}^{2,(t+1)} \\
& \Gamma_l^{(t+1)}\gets \Lambda^{(t+1)}a_l^{(t)} + \sum_\mu F_{\mu l} g_{o,\mu}^{(t+1)}
\end{align*}
 \STATE AMP update of the matrix factorization part \\
\begin{align*}
A^{(t+1)}\gets & \frac{\Delta_I}{N}\sum_\eta\sigma_\eta^{2,(t)} \\
B_\mu^{(t+1)}\gets & \frac{1}{\sqrt N}\sum_\eta S_{\mu\eta}\sigma_\eta^{(t)} \\
&\quad{}-\frac{\Delta_I}{N}\sum_\eta\Sigma_\eta^{(t)}\sigma_\mu^{(t-1)}
\end{align*}
 \STATE AMP update of the estimated marginals $a_l, v_l$ \\
\begin{align*}
& a_l^{(t+1)}\gets f_w(\Lambda^{(t+1)}, \Gamma_l^{(t+1)})  \\
& v_l^{(t+1)}\gets \partial_\Gamma f_w(\Lambda^{(t+1)}, \Gamma_l^{(t+1)})
\end{align*}
 \STATE AMP update of the estimated marginals $\sigma_\nu, \Sigma_\nu$ \\
\begin{align*}
\sigma_\mu^{(t+1)}\gets f_{s,\mu}&(B_\mu^{(t+1)}, A^{(t+1)},\\
&\quad \omega_\mu^{(t+1)}, V^{(t+1)})  \\
\Sigma_\mu^{(t+1)}\gets \partial_Bf_{s,\mu}&(B_\mu^{(t+1)},\\
&\; A^{(t+1)}, \omega_\mu^{(t+1)}, V^{(t+1)})
\end{align*}
 \STATE $t\gets t+1$
 \UNTIL{convergence of $a_l, v_l, \sigma_\mu, \Sigma_\mu$}
 \OUTPUT estimated mean $a_l$ and variance $v_l$ of $w_l$, estimated mean $\sigma_\mu$ and variance $\Sigma_\mu$ of $s_\mu$
\end{algorithmic}
\hrule
\end{multicols}

\subsection{Analysis of the SE equations near the full recovery point}
The state evolution equations are given in section \ref{sec:limiteDense}, eqs.~\eqref{eq:SEdébut}-\eqref{eq:SEfin}. We study the conditions of stability for the fully informative fixed point $(q_s, \hat q_w, q_w) = (1, +\infty, 1)$.

We use the following notation for the update:
\begin{equation}
(q_s, \hat q_w, q_w)^{t+1} = (f_1(r, s), f_2(r, s), f_3(t)) \quad , \quad (r, t, s) = (\Delta_Iq_s^t, \hat q_w^{t+1}, q_w^t)
\end{equation}
where the $f_i$ are given by the SE update equations:
\begin{align}
f_1(r, s) &= \mathbb E_{\xi, \eta} \left[ e^{-r/2} \frac{\left(\sinh(\sqrt r\xi)+\cosh(\sqrt r\xi)\mathrm{erf}\left(\sqrt{\frac{s}{2(1-s)}}\eta\right)\right)^2}{\cosh(\sqrt r\xi)+\sinh(\sqrt r\xi)\mathrm{erf}\left(\sqrt{\frac{s}{2(1-s)}}\eta\right)} \right] \\
f_2(r, s) &= \alpha \mathbb E_{\xi, \eta} \left[ e^{-r/2} \frac{\sinh^2(\sqrt r\xi)\frac{2}{\pi(1-s)}e^{-\eta^2\frac{s}{1-s}}}{\cosh(\sqrt r\xi)+\sinh(\sqrt r\xi)\mathrm{erf}\left(\sqrt{\frac{s}{2(1-s)}}\eta\right)} \right] \\
f_3(t) &= \mathbb E_\xi \left[ e^{-t/2} \sinh(\sqrt t\xi)\tanh(\sqrt t\xi)\right]
\end{align}
where $\xi$ and $\eta$ are standard Gaussians.

We expand around $(1, +\infty, 1)$; we use the parametrization $(r, t, s) = (\Delta_I+\epsilon_r, 1/\epsilon_t^2, 1-\epsilon_s^2)$.

\paragraph*{-- $f_3$ --}
We expand the integrand in $f_3$ around $+\infty$. This is valid only for $\sqrt t\xi=\xi/\epsilon_t\gg 1$; so we introduce a cut-off $\delta$ such that both $\delta = o(1)$ and $\delta = \omega(\epsilon_t)$. For $\xi>\delta$ we use the asymptotic $\sinh(x)\tanh(x) = \frac{1}{2}e^x-\frac{3}{2}e^{-x} + o(e^{-x})$; for $\xi<\delta$ we develop the Gaussian density to the first (constant) order. Then
\begin{align}
f_3(1/\epsilon_t^2) &= e^{-1/2\epsilon_t^2}\, 2\left( \int_0^\infty\frac{\mathrm d\xi}{\sqrt{2\pi}}\,e^{-\xi^2/2} \left(\frac{1}{2}e^{\xi/\epsilon_t}-\frac{3}{2}e^{-\xi/\epsilon_t}\right) - \int_0^\delta\frac{\mathrm d\xi}{\sqrt{2\pi}}\, \left(\frac{1}{2}e^{\xi/\epsilon_t}-\frac{3}{2}e^{-\xi/\epsilon_t}\right) \right) \nonumber \\
 & \quad{}+ e^{-1/2\epsilon_t^2}\, 2\int_0^\delta\frac{\mathrm d\xi}{\sqrt{2\pi}}\, \sinh(\xi/\epsilon_t)\tanh(\xi/\epsilon_t) \\
 &= \frac{1}{2}\left(1+\mathrm{erf}\left(\frac{1}{\sqrt 2\epsilon_t}\right)\right)-\frac{3}{2}\left(1-\mathrm{erf}\left(\frac{1}{\sqrt 2\epsilon_t}\right)\right) \nonumber \\
 & \quad{}- 2e^{-1/2\epsilon_t^2}\epsilon_t\int_0^{\delta/\epsilon_t}\frac{\mathrm d\xi}{\sqrt{2\pi}}\, \left(\frac{1}{2}e^{\xi}-\frac{3}{2}e^{-\xi}\right) +  2e^{-1/2\epsilon_t^2}\epsilon_t\int_0^{\delta/\epsilon_t}\frac{\mathrm d\xi}{\sqrt{2\pi}}\, \sinh(\xi)\tanh(\xi) \\
 &= 1-C_3\epsilon_te^{-1/2\epsilon_t^2} \\
 &\quad\quad C_3 = 2\left(\sqrt{\frac{2}{\pi}} + \int_0^\infty\frac{\mathrm d\xi}{\sqrt{2\pi}}\, \left(\frac{1}{2}e^{\xi}-\frac{3}{2}e^{-\xi} - \sinh(\xi)\tanh(\xi)\right)\right) \approx 1.3
\end{align}
where in the last lines we expanded the error function around $+\infty$.

\paragraph*{-- $f_1$ --}
We use the shorthand notation
\begin{align}
g(x, y) &= \frac{\left(\sinh(x)+\cosh(x)\,\mathrm{erf}(y)\right)^2}{\cosh(x)+\sinh(x)\,\mathrm{erf}(y)} \\
f_1(r, s) &= \mathbb E_{\xi, \eta} \left[ e^{-r/2} g\left(\sqrt r\xi, \eta\sqrt{\frac{s}{2(1-s)}}\right) \right]
\end{align}
Expanding the function is difficult: we can obtain quite easily that
\begin{equation}
f_1(\Delta_I+\epsilon_r, 1-\epsilon_s^2) = 1-C_1(\Delta_I)\epsilon_s+\mathcal O(\epsilon_r\epsilon_s, \ldots)
\end{equation}
but the function $C_1(\Delta_I)$ is harder to obtain. Rather we compute directly the derivative to the constant order:
\begin{align}
\partial_{\epsilon_s} f_1(\Delta_I+\epsilon_r, 1-\epsilon_s^2) &= -C_1(\Delta_I) \\
 &= \mathbb E_{\xi, \eta} \left[ e^{-(\Delta_I+\epsilon_r)/2} \frac{-\eta}{\sqrt 2\epsilon_s^2}\frac{2}{\sqrt\pi}e^{-\eta^2/2\epsilon_s^2} \partial_y g\left(\sqrt{\Delta_I+\epsilon_r}\xi, \frac{\eta}{\sqrt 2\epsilon_s}\right) \right] \\
 &= - \mathbb E_{\xi, \eta} \left[ e^{-(\Delta_I+\epsilon_r)/2} \eta \sqrt{\frac{2}{\pi}} \partial_y g\left(\sqrt{\Delta_I+\epsilon_r}\xi, \frac{\eta}{\sqrt 2}\right) \right] + \mathcal O(\epsilon_s)
\end{align}
So $C_1(\Delta_I) = \sqrt{\frac{2}{\pi}}\mathbb E_{\xi, \eta} \left[ e^{-\Delta_I/2} \eta \partial_y g\left(\sqrt\Delta_I\xi, \eta\right) \right]$.

\paragraph*{-- $f_2$ --}
We introduce
\begin{align}
h(x, y) &= \frac{\sinh^2(x)}{\cosh(x)+\sinh(x)\mathrm{erf}(y/\sqrt 2)} \\
f_2(r,s) &= \alpha \mathbb E_{\xi, \eta} \left[ e^{-r/2}\frac{2}{\pi(1-s)} e^{-\eta^2\frac{s}{1-s}} h\left(\sqrt r\xi, \eta\sqrt{\frac{s}{1-s}}\right) \right]
\end{align}
The first order is enough since it is not constant; we have:
\begin{align}
f_1(\Delta_I+\epsilon_r, 1-\epsilon_s^2) &= \frac{\alpha}{\epsilon_s^2} \mathbb E_{\xi, \eta} \left[ e^{-(\Delta_I+\epsilon_r)/2}\frac{2}{\pi} e^{-\eta^2\frac{1-\epsilon_s^2}{\epsilon_s^2}} h\left(\sqrt{\Delta_I+\epsilon_r}\xi, \eta\sqrt\frac{1-\epsilon_s^2}{\epsilon_s^2}\right) \right] \\
 &= \frac{\alpha}{\epsilon_s} \mathbb E_{\xi, \eta} \left[ e^{-\Delta_I/2}\frac{\sqrt 2}{\pi} h\left(\sqrt\Delta_I\xi, \frac{\eta}{\sqrt 2}\right) + \mathcal O(\epsilon_r, \epsilon_s)\right]
\end{align}
So 
\begin{equation}
f_1(\Delta_I+\epsilon_r, 1-\epsilon_s^2) = \frac{\alpha}{\epsilon_s}C_2(\Delta_I) \quad , \quad C_2(\Delta_I) = \frac{\sqrt 2}{\pi} \mathbb E_{\xi, \eta} \left[ e^{-\Delta_I/2} h\left(\sqrt\Delta_I\xi, \frac{\eta}{\sqrt 2}\right)\right]
\end{equation}
$C_2(\Delta_I)$ is positive for all $\Delta_I$.

\subsubsection{Stability}
We obtain the following update of the perturbation:
\begin{equation}
\begin{pmatrix}
\epsilon_r\\ \epsilon_t\\ \epsilon_s
\end{pmatrix} \to
\begin{pmatrix}
-\Delta_IC_1(\Delta_I)\epsilon_s\\
\sqrt\epsilon_s/\sqrt{\alpha C_2(\Delta_I)}\\
\sqrt C_3\sqrt\epsilon_t e^{-1/4\epsilon_t^2}
\end{pmatrix}
\end{equation}
We consider only the $s$ variable because $r$ does not affect the dynamics and the initialization is done on $r$ and $s$, $t$ being inferred then. We have
\begin{equation}
\label{eq:SEstabilitéEpsilon}
\epsilon_s^{t+1} = \sqrt{C_3}\sqrt[4]{\frac{\epsilon_s^t}{\alpha C_2(\Delta_I)}}e^{-\alpha C_2(\Delta_I)/4\epsilon_s^t}
\end{equation}
which is stable for all $\alpha$ and $\Delta_I$.

Numerically, however, instability can be detected: for $\epsilon_s$ large enough the system diverges from the fully informative fixed point. We compute numerically the limiting $\epsilon_s^*$, such that $\epsilon_s^{t+1}=\epsilon_s^{t}$; we find that $\epsilon_s^*$ tends to zero fast for $\Delta_I$ or $\mu$ going to zero.
\begin{figure}[h!]
 \centering
 \includegraphics[scale=0.65]{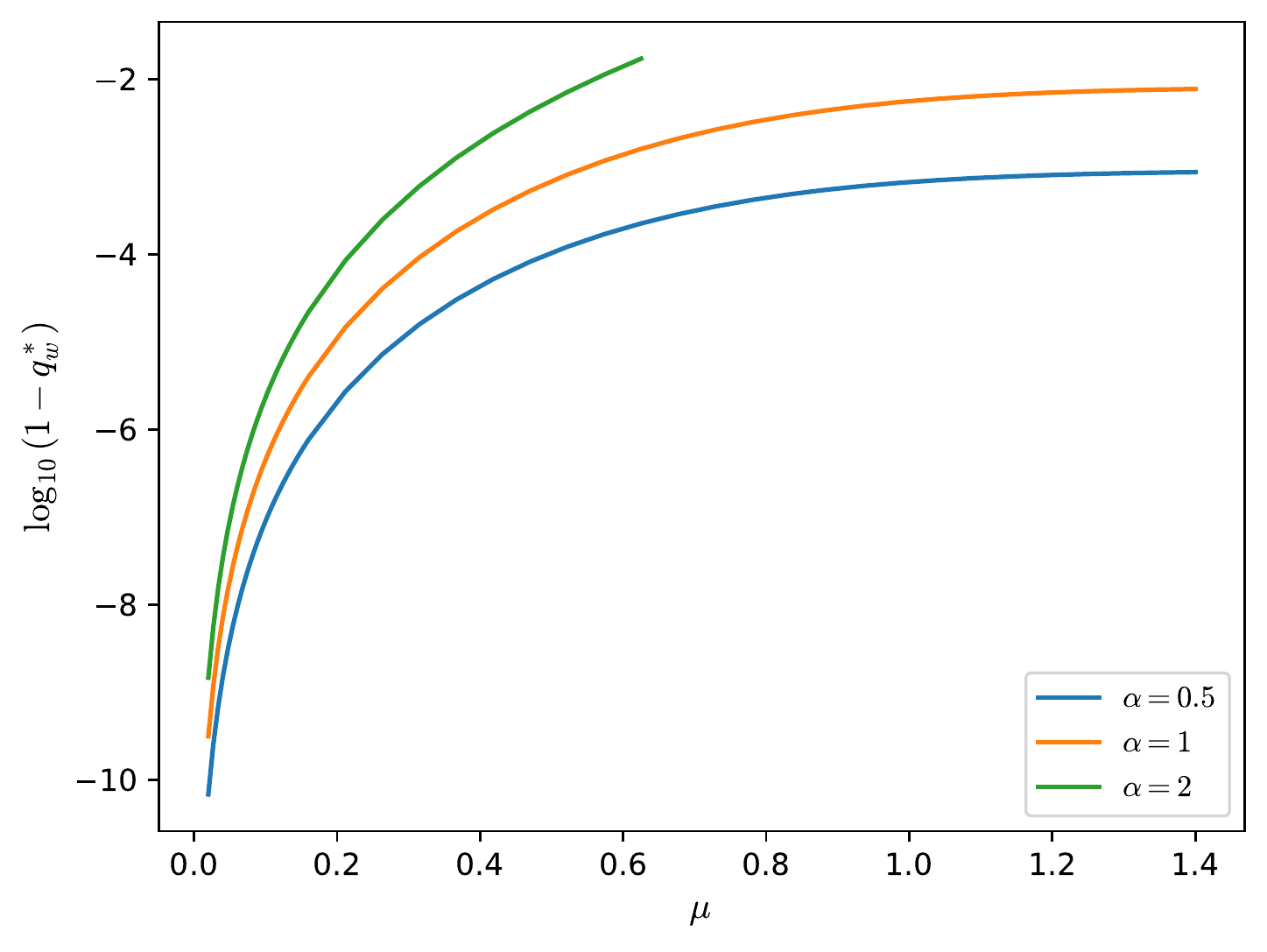}
 \caption{The limiting perturbation $2\log_{10}(\epsilon_s^*)=\log_{10}(1-q_w^*)$ to the fixed point vs $\mu$ for some $\alpha$s. $p_o=1/2$. These are the fixed points of eq. \eqref{eq:SEstabilitéEpsilon}. When initialized above the curves, the system diverges from the fully-informative fixed point. When the snr $\Delta_I=\mu^2/p_o(1-p_o)$ tends to 0, the size of the attraction basin shrinks to 0.}
\end{figure}

In the Gaussian case we have $f_3(t)=t/(1+t)$ and so $\epsilon_s^{t+2} = \epsilon_t^{t+1} = \sqrt{\epsilon_s^t}/\sqrt{\alpha C_2(\Delta_I)}$; so this fixed point is unconditionally not stable.

\subsubsection{Generalization of the prior}
We ask for which prior $P_w$ the fully-informative point is stable. We recall that
\begin{equation}
f_3(t) = \mathbb E_\xi \left[ \left(\int\mathrm dw\, wP_w(w) e^{-tw^2+\sqrt t\xi w}\right)^2\left(\int\mathrm dw\, P_w(w) e^{-tw^2+\sqrt t\xi w}\right)^{-1} \right]
\end{equation}
We assume that $\mathbb E_{P_w}w^2 = \rho_w^2 = 1$.

We show that if $P_w$ admits everywhere a density twice differentiable, then the fully-informative fixed point is unstable. Indeed, at large $t$ we have:
\begin{align}
f_3(t) &= \mathbb E_\xi \left[ \frac{1}{\sqrt t}e^{\frac{1}{2}\xi^2} \left(\int\mathrm dw\, \left(\frac{w}{\sqrt t}+\frac{\xi}{\sqrt t}\right)P_w(\frac{w}{\sqrt t}+\frac{\xi}{\sqrt t}) e^{-\frac{1}{2}w^2} \right)^2  \right. \nonumber \\
 &\quad\quad\quad \left. \left(\int\mathrm dw\, P_w(\frac{w}{\sqrt t}+\frac{\xi}{\sqrt t}) e^{-\frac{1}{2}w^2} \right)^{-1} \right] \\
 &= \frac{1}{\sqrt t}\int\mathrm d\xi\, \left(\int\frac{\mathrm dw}{\sqrt{2\pi}}\, e^{-\frac{1}{2}w^2} \left(\frac{w}{\sqrt t}+\frac{\xi}{\sqrt t}\right) \left(P_w(\frac{\xi}{\sqrt t})+\frac{w}{\sqrt t}P'_w(\frac{\xi}{\sqrt t})\right) \right)^2  \nonumber \\
 &\quad \left(\int\frac{\mathrm dw}{\sqrt{2\pi}}\, e^{-\frac{1}{2}w^2} \left(P_w(\frac{\xi}{\sqrt t})+\frac{w^2}{2t}P''_w(\frac{\xi}{\sqrt t})\right) \right)^{-1} \\
 &= \int\mathrm dx\, \left(xP_w(x)+\frac{1}{t}P'_w(x)\right)^2 \left(P_w(x)+\frac{1}{2t}P''_w(x)\right)^{-1} \\
 &= 1 - \mathcal O(1/t)
\end{align}
and obtain an equation similar to the one of the Gaussian case: $\epsilon_s^{t+2} = C'\sqrt{\epsilon_s^t}$, $C'>0$, which is unstable.

\subsubsection{Large snr}
We give an implicit value for the critical compression ratio $\alpha_\mathrm{algo,d}$. We take the limit $\Delta_I\gg 1$ and seek whether the SE updates converge to the fully informative point.

We can simplify the SE equations to one scalar equation. We expand on $r=\Delta_Iq_s\gg 1$.
We have $q_s=f_1(r,s) \to 1$ for all $s$. As to $q_z$ and $\hat q_z$, we have
\begin{equation}
f_2(s) = \alpha\frac{2}{\pi}\frac{1}{\sqrt{1-s^2}} \mathbb E_\eta \left[1-\mathrm{erf}\left(\eta\sqrt{\frac{s}{2(1+s)}}\right)^2\right]^{-1}
\end{equation}
We plug $f_2$ and $f_3$ together. The fixed points are the $s$ that satisfy the equation $s = f_3(f_2(s))$. The function $f_3\circ f_2$ is plotted in the following figure.
\begin{figure}[ht]
 \centering
 \includegraphics[scale=0.65]{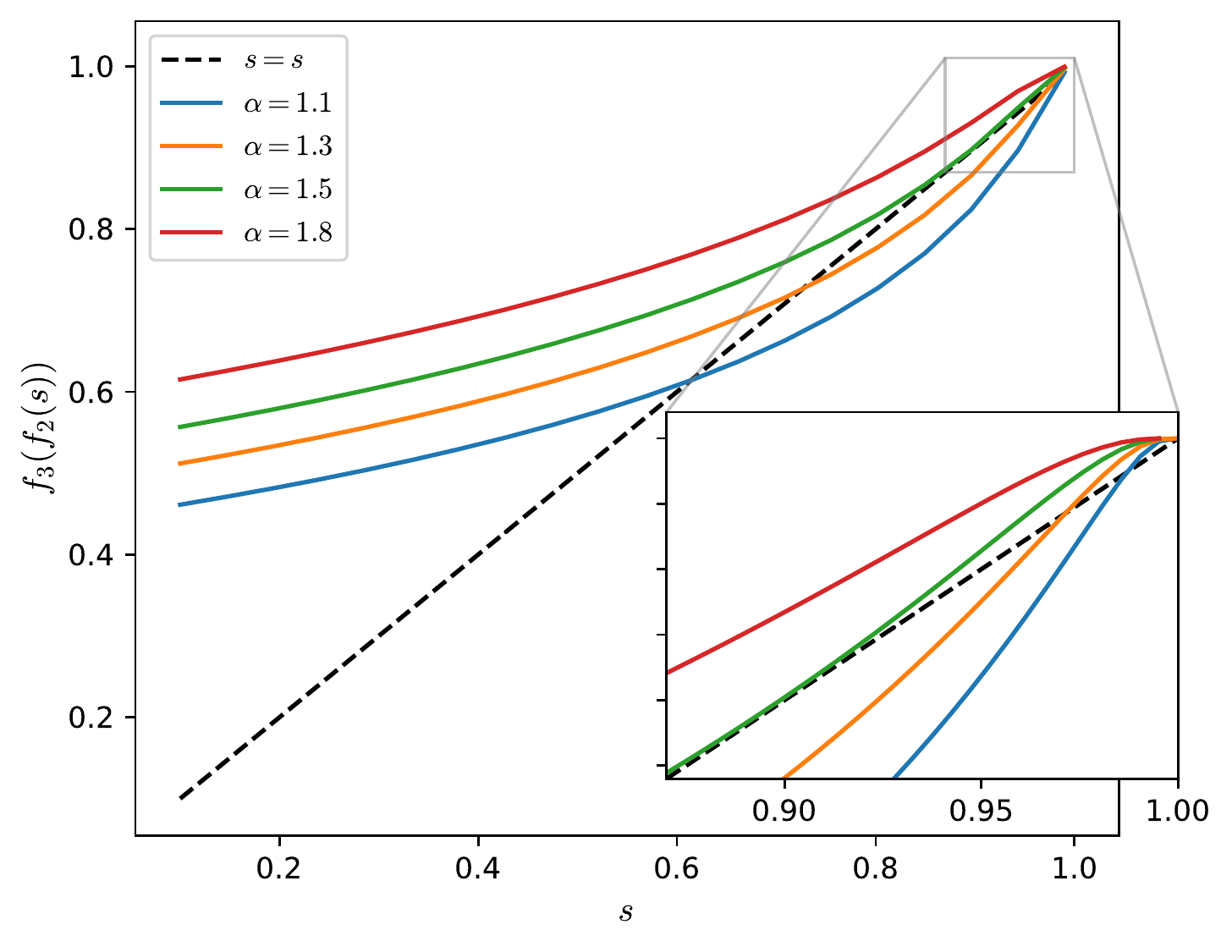}
 \caption{Update function $f_3(f_2(s))$ vs $s$ for many $\alpha$s. It describes the SE equations at large snr. For $\alpha\approx 1.5$ the curve is tangent to the identity at $s\approx 0.9$; for greater $\alpha$ there is only one fixed point, the perfect recovery one.}
\end{figure}

For $\alpha>\alpha_\mathrm{algo,d}\approx 1.5$ the updates lead to perfect recovery $s=1$, starting from any $s$. For $\alpha<\alpha_\mathrm{algo,d}$, perfect recovery is possible only starting from $s$ close to 1; otherwise the iterations lead to a sub-optimal fixed point.

\FloatBarrier
\newpage

\section{Supplementary figures}
\label{sec:figuresSuppl}
\vspace{-2mm}

\begin{figure}[h!]
 \centering
 \includegraphics[width=0.6\linewidth]{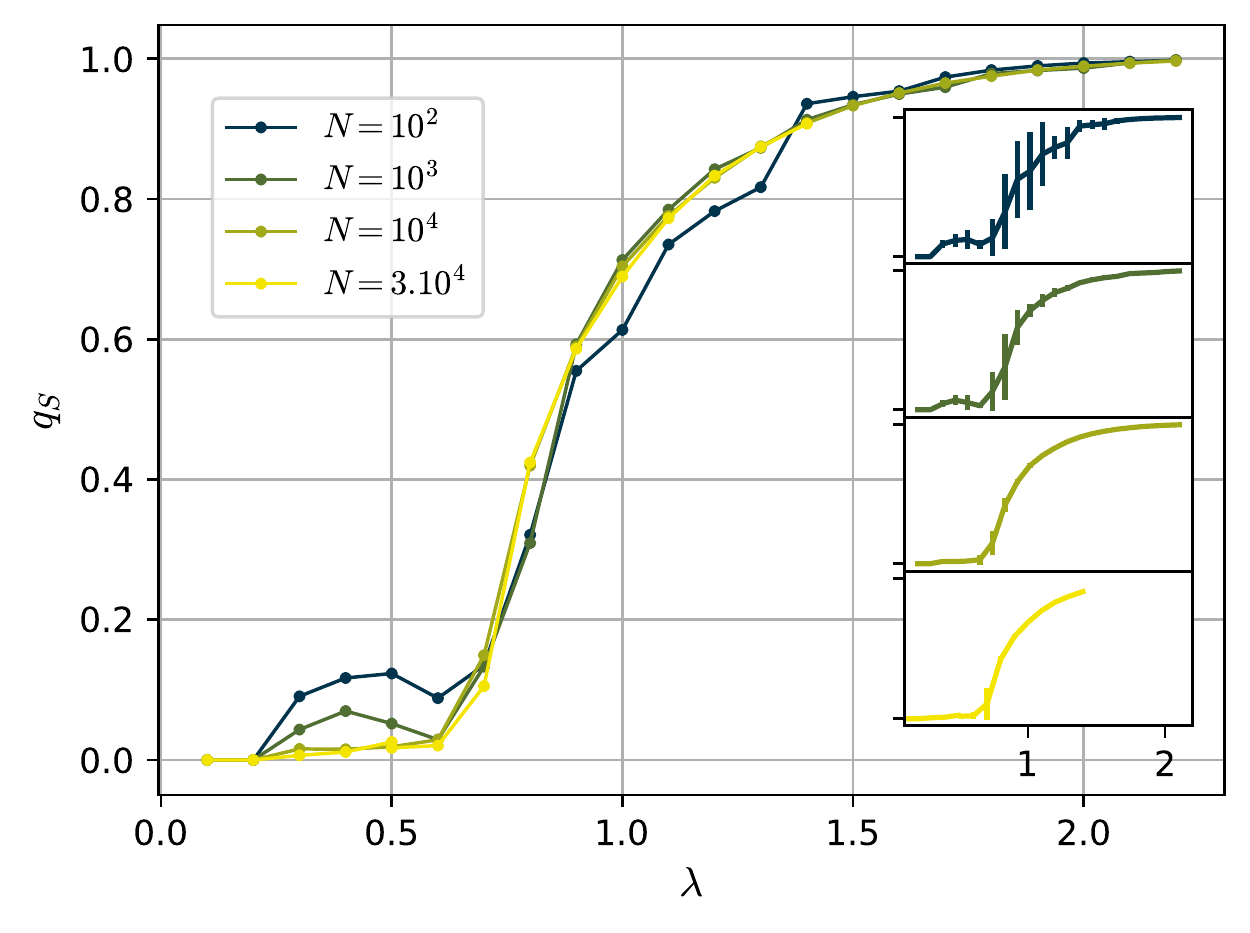}
 \vspace{-4mm}
 \caption{\label{oVsN} Thermodynamic limit. Overlap $q_S$ of the fixed point of the algorithm AMP--BP, vs $\lambda$ for a range of population sizes $N$. $\alpha=3$, $c=5$, $P_w$ Gaussian. We run one hundred ($N$ small) or ten experiments ($N$ large) per point. \emph{Insets:} we plot the standard deviation over the experiments.}
\end{figure}

\vspace{-5mm}

\begin{figure}[h!]
 \centering
 \includegraphics[width=0.5\linewidth]{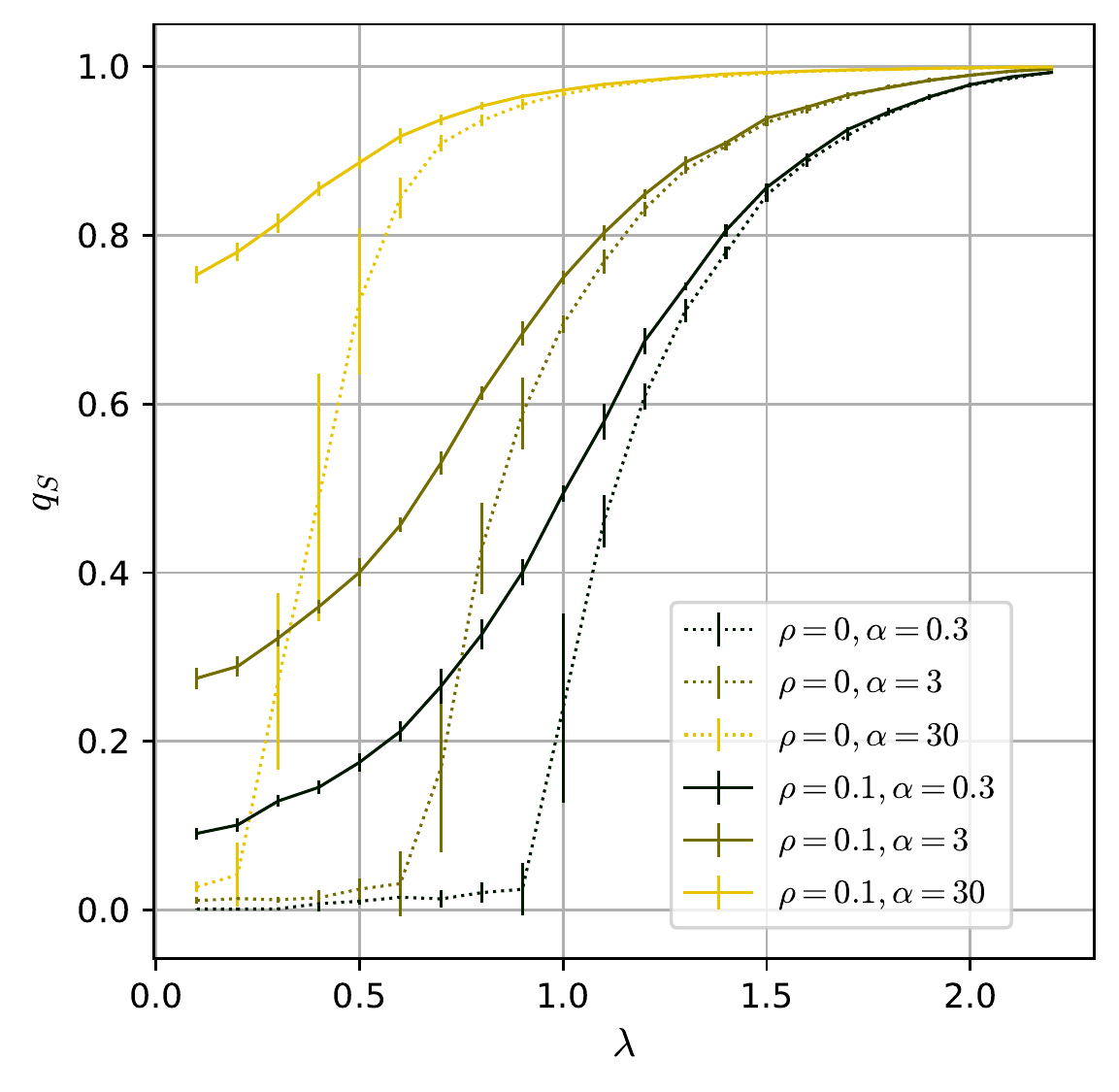}
 \vspace{-4mm}
 \caption{\label{oVsAlphaSS} Semi-supervised. Test overlap $q_S$ of the fixed point of AMP--BP, vs $\lambda$ for a range of compression ratios $\alpha$. The proportion of train nodes is $\rho$. Semi-supervised always performs better than unsupervised. $N=10^4$, $c=5$, $P_w$ Gaussian. We run ten experiments per point.}
\end{figure}

\FloatBarrier

\begin{figure}[h]
 \centering
 \includegraphics[width=0.95\linewidth]{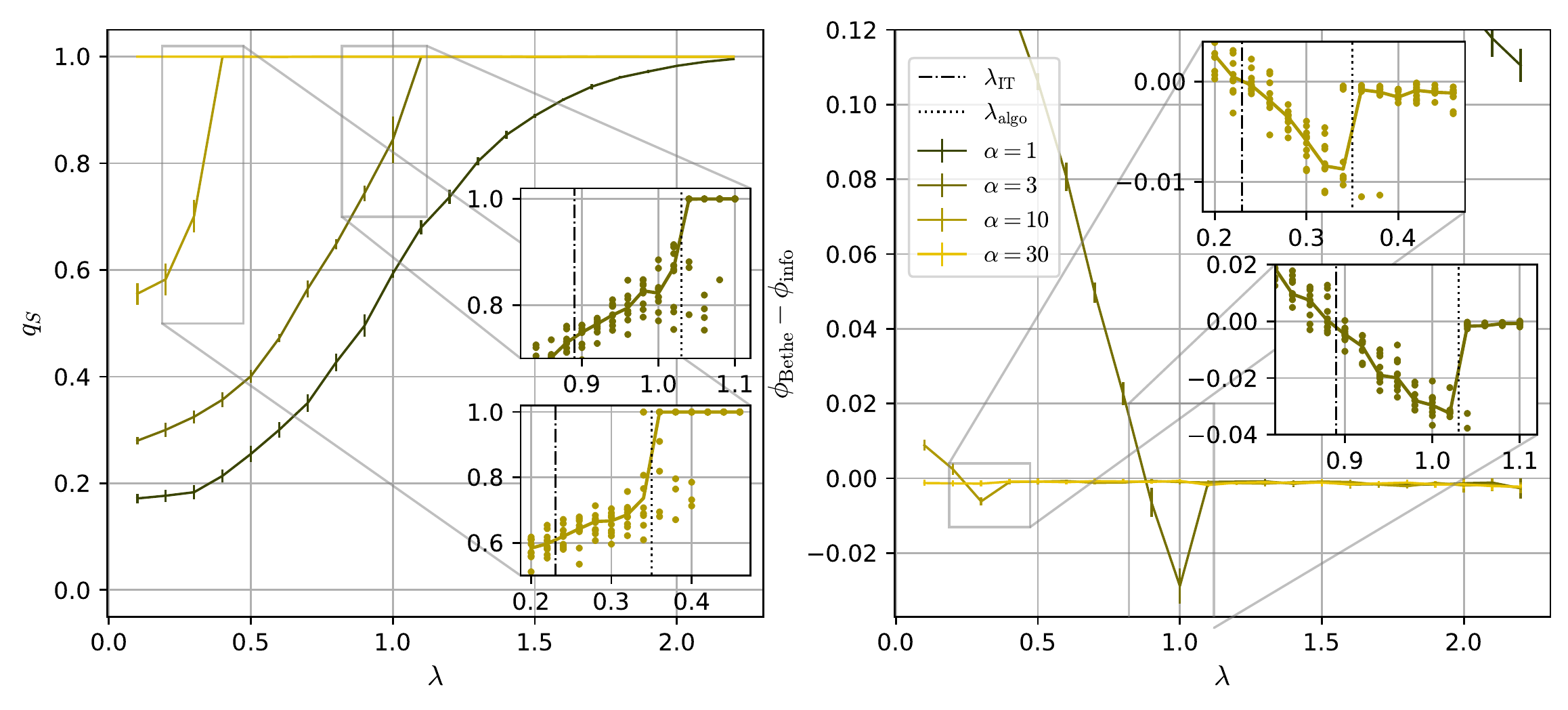}
 \caption{\label{oVsAlphaBinaireSS} Semi-supervised, binary prior. \emph{Left and right:} test overlap $q_S$ and free entropy $\phi_\mathrm{Bethe}-\phi_\mathrm{info}$ of the fixed point of the algorithm AMP--BP, vs $\lambda$ for several compression ratios $\alpha$. $N=10^4$, $c=5$, $P_w$ Rademacher, $\rho=0.1$. We run ten experiments per point; the median is plotted and the error bars are the difference between the 0.85th and 0.15th quantiles. \emph{Insets:} we plot the median and the ten data points. We use damping for AMP--BP: we interpolate taking 1/4 of the values at $t+1$ and 3/4 of the values at~$t$.}
\end{figure}

\begin{figure}[h]
 \centering
 \includegraphics[width=0.48\linewidth]{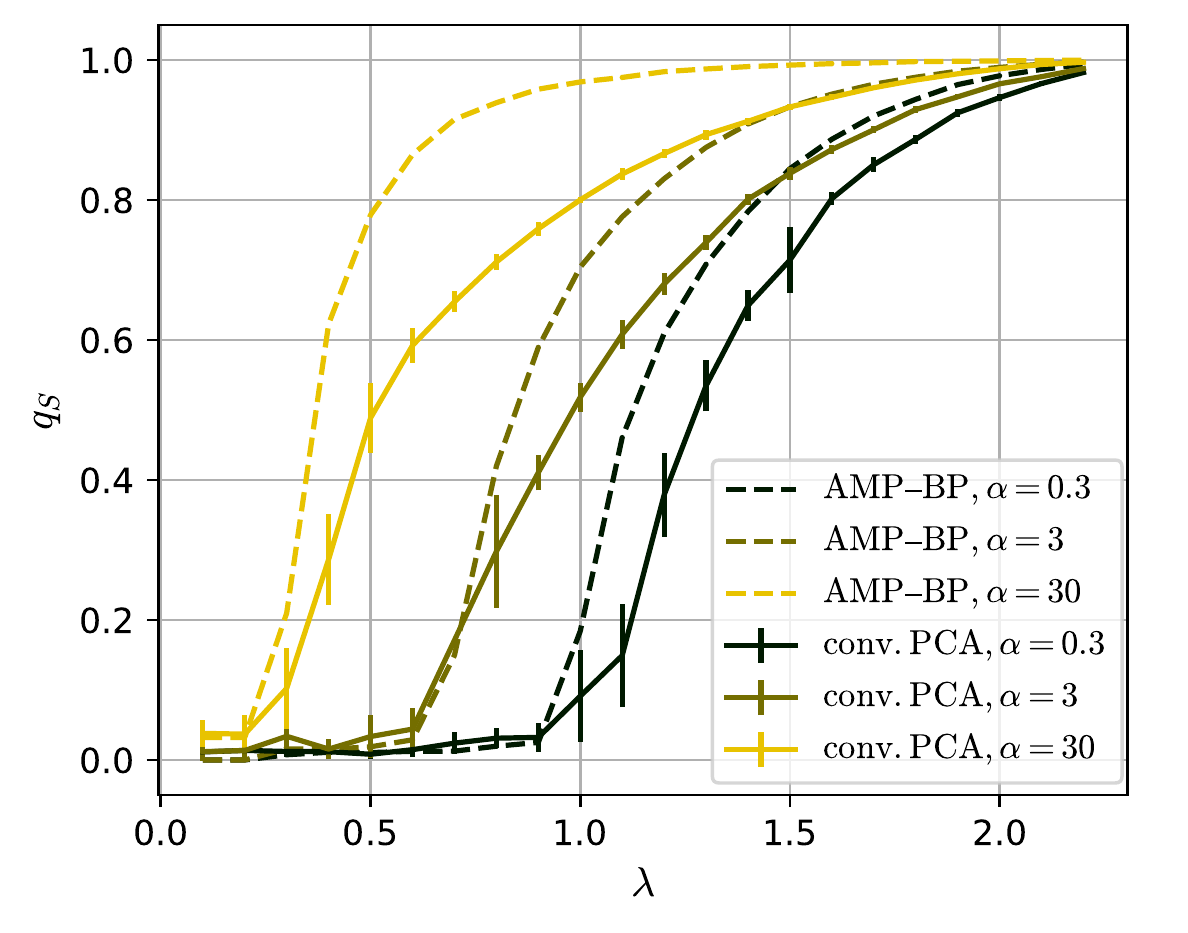}
 \includegraphics[width=0.48\linewidth]{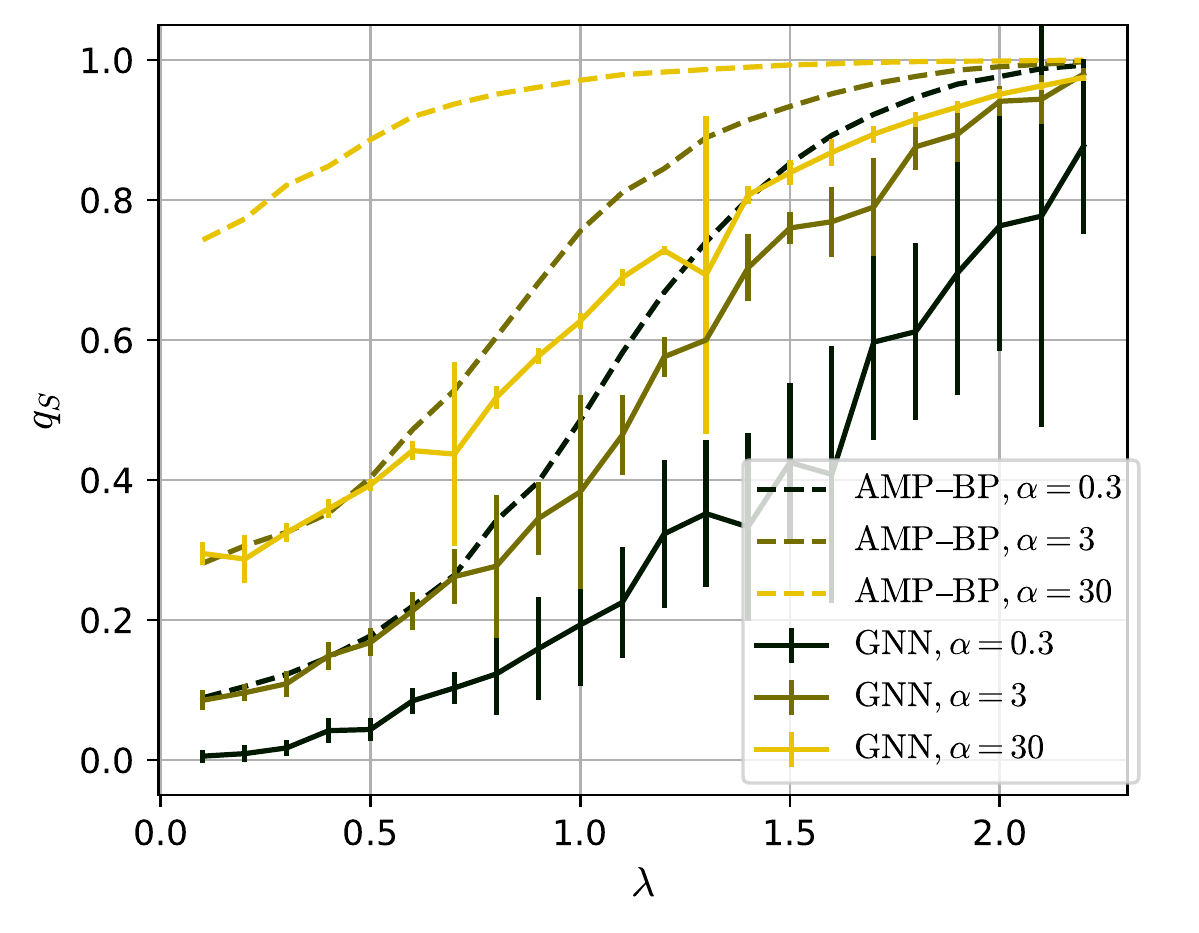}
 \caption{\label{baselineGaussian} Baselines, Gaussian prior. Overlap $q_S$ of the baseline algorithms, vs $\lambda$. We compare to the overlap obtained by AMP--BP. \emph{Left:} unsupervised; for the parameters of the graph convolution we choose $a=0.1$ and $n=4$. \emph{Right:} semi-supervised; for the hyper-parameters of the GNN we choose $n=2$, $N_\mathrm{hidden}=20$, learning rate $3.10^{-4}$ and L2 penalty $10^{-3}$. The train set is $\rho=1/10^\mathrm{th}$ of the nodes. $N=10^4$, $c=5$, $P_w$ Gaussian. We run ten experiments per point.}
\end{figure}

\FloatBarrier

\bibliography{jmlr_23_arxiv}

\begin{thebibliography}{39}
\providecommand{\natexlab}[1]{#1}
\providecommand{\url}[1]{\texttt{#1}}
\expandafter\ifx\csname urlstyle\endcsname\relax
  \providecommand{\doi}[1]{doi: #1}\else
  \providecommand{\doi}{doi: \begingroup \urlstyle{rm}\Url}\fi

\bibitem[Abbe(2017)]{abbe2017community}
Emmanuel Abbe.
\newblock Community detection and stochastic block models: recent developments.
\newblock \emph{The Journal of Machine Learning Research}, 18\penalty0
  (1):\penalty0 6446--6531, 2017.

\bibitem[Abbe et~al.(2015)Abbe, Bandeira, and Hall]{abbe2015exact}
Emmanuel Abbe, Afonso~S Bandeira, and Georgina Hall.
\newblock Exact recovery in the stochastic block model.
\newblock \emph{IEEE Transactions on information theory}, 62\penalty0
  (1):\penalty0 471--487, 2015.

\bibitem[Aubin et~al.(2019)Aubin, Loureiro, Maillard, Krzakala, and
  Zdeborová]{aPrioriGen19}
Benjamin Aubin, Bruno Loureiro, Antoine Maillard, Florent Krzakala, and Lenka
  Zdeborová.
\newblock The spiked matrix model with generative priors.
\newblock In \emph{Advances in Neural Information Processing Systems}, 2019.
\newblock arxiv:1905.12385.

\bibitem[Barbier et~al.(2019)Barbier, Krzakala, Macris, Miolane, and
  Zdeborov{\'a}]{barbier2019optimal}
Jean Barbier, Florent Krzakala, Nicolas Macris, L{\'e}o Miolane, and Lenka
  Zdeborov{\'a}.
\newblock Optimal errors and phase transitions in high-dimensional generalized
  linear models.
\newblock \emph{Proceedings of the National Academy of Sciences}, 116\penalty0
  (12):\penalty0 5451--5460, 2019.

\bibitem[Binkiewicz et~al.(2017)Binkiewicz, Vogelstein, and Rohe]{cSBM14}
Norbert Binkiewicz, Joshua~T. Vogelstein, and Karl Rohe.
\newblock Covariate-assisted spectral clustering.
\newblock \emph{Biometrika}, 104\penalty0 (2):\penalty0 361--377, 2017.
\newblock arxiv:1411.2158.

\bibitem[Celentano et~al.(2021)Celentano, Cheng, and
  Montanari]{celentano2021high}
Michael Celentano, Chen Cheng, and Andrea Montanari.
\newblock The high-dimensional asymptotics of first order methods with random
  data.
\newblock \emph{arXiv preprint arXiv:2112.07572}, 2021.

\bibitem[Cheng et~al.(2022)Cheng, Pan, Hu, and Dokmanić]{GCNRepliques22}
Shi Cheng, Liming Pan, Hong Hu, and Ivan Dokmanić.
\newblock Statistical mechanics of generalization in graph convolution
  networks.
\newblock 2022.
\newblock arxiv:2212.13069.

\bibitem[Chien et~al.(2021)Chien, Peng, Li, and Milenkovic]{benchmarkingCSBM18}
Eli Chien, Jianhao Peng, Pan Li, and Olgica Milenkovic.
\newblock Adaptive universal generalized pagerank graph neural network.
\newblock In \emph{International Conference on Learning Representations}, 2021.
\newblock arxiv:2006.07988.

\bibitem[Cho et~al.(2022)Cho, Min, Kim, Lee, Lee, and Hong]{transformerSBM22}
Sungjun Cho, Seonwoo Min, Jinwoo Kim, Moontae Lee, Honglak Lee, and Seunghoon
  Hong.
\newblock Transformers meet stochastic block models: Attention with
  data-adaptive sparsity and cost.
\newblock In Alice~H. Oh, Alekh Agarwal, Danielle Belgrave, and Kyunghyun Cho,
  editors, \emph{Advances in Neural Information Processing Systems}, 2022.
\newblock URL \url{https://openreview.net/forum?id=w_jvWzNXd6n}.

\bibitem[Coja-Oghlan et~al.(2017)Coja-Oghlan, Krzakala, Perkins, and
  Zdeborov{\'a}]{coja2017information}
Amin Coja-Oghlan, Florent Krzakala, Will Perkins, and Lenka Zdeborov{\'a}.
\newblock Information-theoretic thresholds from the cavity method.
\newblock In \emph{Proceedings of the 49th Annual ACM SIGACT Symposium on
  Theory of Computing}, pages 146--157, 2017.

\bibitem[Cornacchia et~al.(2022)Cornacchia, Mignacco, Veiga, Gerbelot,
  Loureiro, and Zdeborov{\'a}]{cornacchia2022learning}
Elisabetta Cornacchia, Francesca Mignacco, Rodrigo Veiga, C{\'e}dric Gerbelot,
  Bruno Loureiro, and Lenka Zdeborov{\'a}.
\newblock Learning curves for the multi-class teacher-student perceptron.
\newblock \emph{Machine Learning: Science and Technology}, 2022.

\bibitem[Decelle et~al.(2011{\natexlab{a}})Decelle, Krzakala, Moore, and
  Zdeborov{\'a}]{decelle2011inference}
Aurelien Decelle, Florent Krzakala, Cristopher Moore, and Lenka Zdeborov{\'a}.
\newblock Inference and phase transitions in the detection of modules in sparse
  networks.
\newblock \emph{Physical Review Letters}, 107\penalty0 (6):\penalty0 065701,
  2011{\natexlab{a}}.

\bibitem[Decelle et~al.(2011{\natexlab{b}})Decelle, Krzakala, Moore, and
  Zdeborová]{SBM11}
Aurélien Decelle, Florent Krzakala, Cristopher Moore, and Lenka Zdeborová.
\newblock Asymptotic analysis of the stochastic block model for modular
  networks and its algorithmic applications.
\newblock \emph{Phys. Rev. E}, 84, 2011{\natexlab{b}}.
\newblock arxiv:1109.3041.

\bibitem[Deshpande et~al.(2018)Deshpande, Sen, Montanari, and Mossel]{cSBM18}
Yash Deshpande, Subhabrata Sen, Andrea Montanari, and Elchanan Mossel.
\newblock Contextual stochastic block models.
\newblock In S.~Bengio, H.~Wallach, H.~Larochelle, K.~Grauman, N.~Cesa-Bianchi,
  and R.~Garnett, editors, \emph{Advances in Neural Information Processing
  Systems}, volume~31, 2018.
\newblock arxiv:1807.09596.

\bibitem[Donoho et~al.(2009)Donoho, Maleki, and Montanari]{donoho2009message}
David~L Donoho, Arian Maleki, and Andrea Montanari.
\newblock Message-passing algorithms for compressed sensing.
\newblock \emph{Proceedings of the National Academy of Sciences}, 106\penalty0
  (45):\penalty0 18914--18919, 2009.

\bibitem[Fortunato(2010)]{fortunato2010community}
Santo Fortunato.
\newblock Community detection in graphs.
\newblock \emph{Physics reports}, 486\penalty0 (3-5):\penalty0 75--174, 2010.

\bibitem[Fountoulakis et~al.(2022)Fountoulakis, He, Lattanzi, Perozzi,
  Tsitsulin, and Yang]{attentionCSBM22}
Kimon Fountoulakis, Dake He, Silvio Lattanzi, Bryan Perozzi, Anton Tsitsulin,
  and Shenghao Yang.
\newblock On classification thresholds for graph attention with edge features.
\newblock arxiv:2210.10014, 2022.

\bibitem[Gabri{\'e} et~al.(2018)Gabri{\'e}, Manoel, Luneau, Macris, Krzakala,
  Zdeborov{\'a}, et~al.]{gabrie2018entropy}
Marylou Gabri{\'e}, Andre Manoel, Cl{\'e}ment Luneau, Nicolas Macris, Florent
  Krzakala, Lenka Zdeborov{\'a}, et~al.
\newblock Entropy and mutual information in models of deep neural networks.
\newblock \emph{Advances in Neural Information Processing Systems}, 31, 2018.

\bibitem[Gamarnik et~al.(2022)Gamarnik, Moore, and
  Zdeborov{\'a}]{gamarnik2022disordered}
David Gamarnik, Cristopher Moore, and Lenka Zdeborov{\'a}.
\newblock Disordered systems insights on computational hardness.
\newblock \emph{Journal of Statistical Mechanics: Theory and Experiment},
  2022\penalty0 (11):\penalty0 114015, 2022.

\bibitem[Gerbelot and Berthier(2021)]{gerbelot2021graph}
C{\'e}dric Gerbelot and Rapha{\"e}l Berthier.
\newblock Graph-based approximate message passing iterations.
\newblock \emph{arXiv preprint arXiv:2109.11905}, 2021.

\bibitem[Gy{\"o}rgyi(1990)]{gyorgyi1990first}
G{\'e}za Gy{\"o}rgyi.
\newblock First-order transition to perfect generalization in a neural network
  with binary synapses.
\newblock \emph{Physical Review A}, 41\penalty0 (12):\penalty0 7097, 1990.

\bibitem[Krzakala et~al.(2012)Krzakala, Mézard, Sausset, Sun, and
  Zdeborová]{GLM12}
Florent Krzakala, Marc Mézard, Francois Sausset, Yifan Sun, and Lenka
  Zdeborová.
\newblock Probabilistic reconstruction in compressed sensing: algorithms, phase
  diagrams, and threshold achieving matrices.
\newblock \emph{Journal of Statistical Mechanics: Theory and Experiment},
  2012\penalty0 (08), aug 2012.
\newblock arxiv:1206.3953.

\bibitem[Krzakala et~al.(2013)Krzakala, Moore, Mossel, Neeman, Sly, Zdeborová,
  and Zhang]{SBM13spectral}
Florent Krzakala, Cristopher Moore, Elchanan Mossel, Joe Neeman, Allan Sly,
  Lenka Zdeborová, and Pan Zhang.
\newblock Spectral redemption in clustering sparse networks.
\newblock \emph{Proceedings of the National Academy of Sciences}, 110(52),
  2013.

\bibitem[Krzakala et~al.(2014)Krzakala, Manoel, Tramel, and
  Zdeborová]{freeEntropyGLM14}
Florent Krzakala, Andre Manoel, Eric~W. Tramel, and Lenka Zdeborová.
\newblock Variational free energies for compressed sensing.
\newblock In \emph{2014 IEEE International Symposium on Information Theory},
  pages 1499--1503, 2014.
\newblock arxiv:1402.1384.

\bibitem[Lesieur et~al.(2017)Lesieur, Krzakala, and
  Zdeborov{\'a}]{lesieur2017constrained}
Thibault Lesieur, Florent Krzakala, and Lenka Zdeborov{\'a}.
\newblock Constrained low-rank matrix estimation: Phase transitions,
  approximate message passing and applications.
\newblock \emph{Journal of Statistical Mechanics: Theory and Experiment},
  2017\penalty0 (7):\penalty0 073403, 2017.

\bibitem[Loureiro et~al.(2021)Loureiro, Gerbelot, Cui, Goldt, Krzakala, Mezard,
  and Zdeborov{\'a}]{loureiro2021learning}
Bruno Loureiro, Cedric Gerbelot, Hugo Cui, Sebastian Goldt, Florent Krzakala,
  Marc Mezard, and Lenka Zdeborov{\'a}.
\newblock Learning curves of generic features maps for realistic datasets with
  a teacher-student model.
\newblock \emph{Advances in Neural Information Processing Systems},
  34:\penalty0 18137--18151, 2021.

\bibitem[Lu and Sen(2020)]{lu2020contextual}
Chen Lu and Subhabrata Sen.
\newblock Contextual stochastic block model: Sharp thresholds and contiguity.
\newblock \emph{arXiv preprint arXiv:2011.09841}, 2020.

\bibitem[Manoel et~al.(2017)Manoel, Krzakala, M{\'e}zard, and
  Zdeborov{\'a}]{manoel2017multi}
Andre Manoel, Florent Krzakala, Marc M{\'e}zard, and Lenka Zdeborov{\'a}.
\newblock Multi-layer generalized linear estimation.
\newblock In \emph{2017 IEEE International Symposium on Information Theory
  (ISIT)}, pages 2098--2102. IEEE, 2017.

\bibitem[Miolane(2017)]{miolane2017fundamental}
L{\'e}o Miolane.
\newblock Fundamental limits of low-rank matrix estimation: the non-symmetric
  case.
\newblock \emph{arXiv preprint arXiv:1702.00473}, 2017.

\bibitem[Mossel et~al.(2015)Mossel, Neeman, and Sly]{mossel2015reconstruction}
Elchanan Mossel, Joe Neeman, and Allan Sly.
\newblock Reconstruction and estimation in the planted partition model.
\newblock \emph{Probability Theory and Related Fields}, 162:\penalty0 431--461,
  2015.

\bibitem[Mossel et~al.(2018)Mossel, Neeman, and Sly]{mossel2018proof}
Elchanan Mossel, Joe Neeman, and Allan Sly.
\newblock A proof of the block model threshold conjecture.
\newblock \emph{Combinatorica}, 38\penalty0 (3):\penalty0 665--708, 2018.

\bibitem[Ongie et~al.(2020)Ongie, Jalal, Metzler, Baraniuk, Dimakis, and
  Willett]{ongie2020deep}
Gregory Ongie, Ajil Jalal, Christopher~A Metzler, Richard~G Baraniuk,
  Alexandros~G Dimakis, and Rebecca Willett.
\newblock Deep learning techniques for inverse problems in imaging.
\newblock \emph{IEEE Journal on Selected Areas in Information Theory},
  1\penalty0 (1):\penalty0 39--56, 2020.

\bibitem[Peixoto(2019)]{peixoto2019bayesian}
Tiago~P Peixoto.
\newblock Bayesian stochastic blockmodeling.
\newblock \emph{Advances in network clustering and blockmodeling}, pages
  289--332, 2019.

\bibitem[Shlezinger et~al.(2020)Shlezinger, Whang, Eldar, and
  Dimakis]{shlezinger2020model}
Nir Shlezinger, Jay Whang, Yonina~C Eldar, and Alexandros~G Dimakis.
\newblock Model-based deep learning.
\newblock \emph{arXiv preprint arXiv:2012.08405}, 2020.

\bibitem[Sompolinsky et~al.(1990)Sompolinsky, Tishby, and
  Seung]{sompolinsky1990learning}
Haim Sompolinsky, Naftali Tishby, and H~Sebastian Seung.
\newblock Learning from examples in large neural networks.
\newblock \emph{Physical Review Letters}, 65\penalty0 (13):\penalty0 1683,
  1990.

\bibitem[Tsitsulin et~al.(2021)Tsitsulin, Rozemberczki, Palowitch, and
  Perozzi]{benchmarkingCSBM22}
Anton Tsitsulin, Benedek Rozemberczki, John Palowitch, and Bryan Perozzi.
\newblock Synthetic graph generation to benchmark graph learning.
\newblock In \emph{Workshop on Graph Learning Benchmarks}, 2021.
\newblock arxiv:2204.01378.

\bibitem[Yang et~al.(2013)Yang, McAuley, and Leskovec]{yang2013community}
Jaewon Yang, Julian McAuley, and Jure Leskovec.
\newblock Community detection in networks with node attributes.
\newblock In \emph{2013 IEEE 13th international conference on data mining},
  pages 1151--1156. IEEE, 2013.

\bibitem[Zdeborov{\'a} and Krzakala(2016)]{inferences15}
Lenka Zdeborov{\'a} and Florent Krzakala.
\newblock Statistical physics of inference: Thresholds and algorithms.
\newblock \emph{Advances in Physics}, 65\penalty0 (5):\penalty0 453--552, 2016.

\bibitem[Zhang et~al.(2014)Zhang, Moore, and Zdeborov{\'a}]{zhang2014phase}
Pan Zhang, Cristopher Moore, and Lenka Zdeborov{\'a}.
\newblock Phase transitions in semisupervised clustering of sparse networks.
\newblock \emph{Physical Review E}, 90\penalty0 (5):\penalty0 052802, 2014.

\end{thebibliography}

\end{document}